\documentclass{article} 
\usepackage{jcappub}
\usepackage{epstopdf}
\usepackage{graphicx}
\usepackage{color}
\usepackage{hyperref}

\newcommand{\aap}{Astron.\ Astrophys.\ }
\newcommand{\aj}{Astron.\ J.\ }
\newcommand{\apj}{Astrophys.\ J.\  }

\newcommand{\mnras}{Mon.\ Not.\ R.\ Astron.\ Soc.\ }
 
\newcommand{\physrep}{{\em Phys.\ Rep.}}

\newcommand{\prd}{Phys.\ Rev.\ D\ }
\newcommand{\jcap}{Journal of Cosmology and Astro-Particle Physics}

\newcommand{\bdm}{\begin{displaymath}}
\newcommand{\edm}{\end{displaymath}}

\newcommand{\agt}{\gtrsim}

\newcommand{\bea}{\begin{eqnarray}}
\newcommand{\eea}{\end{eqnarray}}

\newcommand{\be}{\begin{equation}}
\newcommand{\ee}{\end{equation}}
\newcommand{\lp}{\left(}
\newcommand{\rp}{\right)}
\newcommand{\lb}{\left[}
\newcommand{\rb}{\right]}
\newcommand{\beal}{\begin{align}}
\newcommand{\eal}{\end{align}}

\newcommand{\HH}{{\cal H}}

\newcommand{\vp}{\varphi}
\newcommand{\bk}{{\vec k}}
\def\ds{\displaystyle}

\def\Mpc{\, h^{-1} \, {\rm Mpc}}

\def\kMpc{\, h \, {\rm Mpc}^{-1}}

\title{Nonlinear effects of dark energy clustering beyond the acoustic scales}

\author[a,b]{Stefano Anselmi,}
\author[b,c]{Diana L{\' o}pez Nacir,}
\author[d,b]{and Emiliano Sefusatti}

\affiliation[a]{Department of Physics/CERCA/ISO, Case Western Reserve University, Cleveland, OH 44106-7079 -- USA}
\affiliation[b]{The Abdus Salam International Center for Theoretical Physics, strada costiera 11, I-34151 Trieste -- Italy}
\affiliation[c]{Departamento de F\'\i sica and IFIBA, FCEyN UBA, Facultad de Ciencias Exactas y Naturales,
 Ciudad Universitaria, Pabell\' on I, 1428 Buenos Aires  --  Argentina}
\affiliation[d]{INAF - Osservatorio Astronomico di Brera, via E. Bianchi 46, I-23807 Merate (LC) -- Italy}

\emailAdd{stefano.anselmi@case.edu}
\emailAdd{dlopez\_n@ictp.it}
\emailAdd{emiliano.sefusatti@brera.inaf.it}

\abstract{We extend the resummation method of Anselmi \& Pietroni (2012) to compute the total density power spectrum in models of quintessence characterized by a vanishing speed of sound. For standard $\Lambda$CDM cosmologies, this resummation scheme allows predictions with an accuracy at the few percent level beyond the range of scales where acoustic oscillations are present, therefore comparable to other, common numerical tools. In addition, our theoretical approach indicates an approximate but valuable and simple relation between the power spectra for standard quintessence models and models where scalar field perturbations appear at all scales. This, in turn, provides an educated guess for the prediction of nonlinear growth in models with generic speed of sound, particularly valuable since no numerical results are yet available.}


\begin{document}
\maketitle
\flushbottom

%%%%%%%%%%%%%%%%%%%%%%%%%%%%%%%%%%%%%%%%%%%%%%%%%%%%%%%%%%%%%%%%%%%%%%%%%%%%%%%%%%%%%%%%%%%%%%
%%%%%%%%%%%%%%%%%%%%%%%%%%%%%%%%%%%%%%%%%%%%%%%%%%%%%%%%%%%%%%%%%%%%%%%%%%%%%%%%%%%%%%%%%%%%%%
\section{Introduction}

Ever since the discovery of the cosmic acceleration found by the observations of type Ia supernovae in 1998 \cite{RiessEtal1998, PerlmutterEtal1999}, a tremendous amount of observational and theoretical effort has been put in place in order to understand the cause of this phenomenon. 

A plethora of alternative models attribute the observed acceleration to either the presence of a mysterious energy component dubbed dark energy (DE) which dominates at the current epoch, or to  a modification of the gravitational laws on cosmological distances (see, e.g. \cite{CopelandSamiTsujikawa2006} for an overview of proposed theoretical models). While the standard $\Lambda$CDM model, where the acceleration is driven by a cosmological constant $\Lambda$, provides a good fit to observations, several models cannot, so far, be ruled-out. The next generation of large-scale structure (LSS) surveys such as EUCLID \cite{LaureijsEtal2011} and DESI\footnote{\href{http://desi.lbl.gov/}{http://desi.lbl.gov/}} will provide an unprecedented determination of observables such as the weak lensing shear and galaxy power spectra over a very large range of scales, and will be therefore sensitive to the details of structure formation, representing a unique opportunity to discriminate between competing models. However, the 
improvement in the quality and 
quantity of observational data represents a challenge to the accuracy of theoretical predictions, particularly for non-standard models. 

A fundamental complication, in this respect, is given by the nonlinear nature of structure formation. Indeed many cosmological probes of matter density correlations will present small statistical errors on scales smaller than $\sim 100\Mpc$, where gravitational instability is responsible for relevant nonlinear corrections to linear theory predictions. The solution to the problem entails two goals: to understand and predict these corrections on one hand and, on the other, to achive a fast way (seconds) to compute numerically such predictions, in order to efficiently sample the theory parameter space in the data analysis. Our current understanding of the nonlinear regime, rather limited even for the standard paradigm, is very poor for non--standard cosmological models. In the following we will present an important step in this direction for the case of models with perturbations in the quintessence field.

Quintessence is one of the most popular models of DE, where the acceleration of the Universe is attributed to a scalar field with negative pressure \cite{ZlatevWangSteinhardt1999}. Its standard formulation involves a minimally coupled canonical scalar field, whose fluctuations propagate at the speed of light $c_s=1$. In this case, sound waves maintain the quintessence homogeneous on scales smaller than the Hubble radius $H^{-1}$ \cite{FerreiraJoyce1997}. The quintessence contribution to the total energy density affects the  expansion of the Universe and the growth of dark matter perturbations, allowing observational constraints on the DE equation of state $w(t)=p_Q(t)/\rho_Q(t)$, where $p_Q$ and $\rho_Q$ are the pressure and energy density of the scalar field. In this set-up, quintessence clustering effects take place only on scales of order $H^{-1}$, and their detection is therefore severely limited by cosmic variance. However, this may not be the case for more general quintessence models where the scalar 
field kinetic term is non-canonical\footnote{ The name ``quintessence'' is usually reserved for theories where the dark energy is given by a scalar field with a canonical kinetic term, however, following previous works on the subject, we will refer as quintessence to a generic minimally coupled scalar field, with either canonical or non-canonical kinetic term.}.  In fact,  in \cite{CreminelliEtal2006B,CreminelliEtal2009}, following the approach of \cite{CheungEtal2008},  the authors constructed the most general effective field theory  describing quintessence perturbations around a given Friedman-Robertson-Walker background space-time, showing the existence of theoretically consistent models where the fluctuations propagate at a speed $c_s\neq 1$.   In particular, by a careful analysis of the theoretical consistency of the models, namely, the classical stability of the perturbations and the absence of ghosts, ref.~\cite{CreminelliEtal2009} concluded that the region of parameter space with $w<-1$ (the so-
called 
``phantom regime'') is in general  excluded, unless the  fluctuations are characterized by a practically  vanishing, imaginary speed of sound, e.g. $|c_s|\lesssim 10^{-15}$ (note that observational limits  do not exclude  the case $w<-1$  \cite{Planck2013parameters,RestEtal2013}).
 In the latter case,   the stability of the perturbations would be guaranteed by the presence of higher derivative operators with negligible effects on cosmological scales. As was emphasized in  \cite{CreminelliEtal2006B,CreminelliEtal2009}, such a tiny speed of sound should not be considered as a fine tuning, since in the limit $c_s\to 0$ one recovers the  ghost condensate model of ref.~\cite{ArkaniHamedEtal2004A}, which is invariant under a shift symmetry of the field,  and hence  a value of $c_s\simeq 0$ can be interpreted as a deformation of the limit where such symmetry is recovered.
  
What makes these models with $c_s\simeq 0$ of particular interest is that quintessence perturbations can cluster on all observable scales, opening up to a rich phenomenology. In fact, for such models the quintessence field is expected to affect not only the growth history of dark matter through the background evolution but also by actively participating to structure formation, both at the linear and nonlinear level. In the linear regime, the observational consequences of clustering quintessence have been investigated by several authors: see for instance \cite{WellerLewis2003, BeanDore2004, DeDeoCaldwellSteinhardt2003, Hannestad2005, EricksonEtal2002, DePutterHutererLinder2010, SaponeKunz2009} for large scale CMB anisotropies, \cite{Takada2006} for galaxy redshift surveys, \cite{TorresRodriguezCress2007} for neutral hydrogen observations or \cite{HuScranton2004, CorasanitiGiannantonioMelchiorri2005} for the cross-correlation of the Integrated Sachs-Wolfe effect in the CMB and the LSS. However, the possibility 
to detect 
quintessence perturbations effects focusing on observations in the linear regime alone is limited by large degeneracies with cosmological parameters, particularly when quintessence clusters on all relevant scales. In this case, in fact, such effects consist essentially in a change of the fluctuations amplitude. On the other hand the relation between the linear {\em and} nonlinear growth of density perturbations can be affected by clustering quintessence in a peculiar way, leading to specific signatures that could in principle allow to distinguish these models from $\Lambda$CDM and homogeneous quintessence cosmologies \cite{SefusattiVernizzi2011, DAmicoSefusatti2011, AnselmiBallesterosPietroni2011}.

Robust predictions of the nonlinear dynamics of the structure formation process ultimately require numerical simulations. The case of homogeneous quintessence (i.e., $c_s=1$), as it affects only the background evolution has been investigated by means of N-body simulations in several articles (see for instance \cite{Baldi2012, KuhlenVogelsbergerAngulo2012} and references therein). The case of clustering quintessence, however, represents a much more difficult problem and numerical results are still lacking. The extension of common fitting formulas such as \texttt{halofit} \cite{SmithEtal2003, TakahashiEtal2012, Zhao2013}, based in turn on numerical simulations, is also, {\em a priori} not possible. Precisely for this reason, the study of analytical approximations in cosmological perturbation theory (PT) to investigate the mildly nonlinear regime is particularly relevant. On one hand it can give insights in the physical processes at play. One of the main results of this work, in fact, consists in showing how 
fitting functions valid for tested, standard cosmologies can be used in clustering quintessence models. Such extensions have been already considered in the literature without a proper theoretical justification \cite{SaponeKunz2009, AmendolaKunzSapone2008, SaponeKunzAmendola2010}. On the other hand, PT predictions can as well provide useful checks for future numerical results, which are likely to comprise significant approximations in their description of the dynamics of quintessence perturbations.
 
The nonlinear regime of structure formation in quintessence models with $c_s=0$ has been studied in previous literature. Several works focussed on the spherical collapse of structures in the presence of quintessence perturbations \cite{MotaVanDeBruck2004, NunesMota2006, AbramoEtal2007}. In addition, ref.~\cite{CreminelliEtal2010} makes use of the spherical collapse model \cite{GunnGott1972} (see \cite{BasseEggersBjaeldeWong2011, BatistaPace2013} for the case of arbitrary sound speed) along with the Press-Schechter formalism \cite{PressSchechter1974}, to provide predictions for the halo mass function. We will consider here, instead, analytical predictions in cosmological perturbation theory for the mildly nonlinear regime of the density power spectrum (see \cite{BernardeauEtal2002} for a review of standard PT).  

In the context of standard, $\Lambda$CDM cosmologies, perturbative techniques have witnessed a significant progress, motivated by the need for accurate predictions for Baryon Acoustic Oscillations measurements, following the seminal papers of Crocce \& Scoccimarro \cite{CrocceScoccimarro2006A, CrocceScoccimarro2006B, CrocceScoccimarro2008}. Several different ÒresummedÓ perturbative schemes and approaches have been proposed in the last few years \cite{MatarresePietroni2007, TaruyaHiramatsu2008, Pietroni2008, BernardeauCrocceScoccimarro2008, BernardeauVanDeRijtVernizzi2012, JurgensBartelmann2012, BlasGarnyKonstandin2014}, along with public codes implementing some of the methods \cite{CrocceScoccimarroBernardeau2012,TaruyaEtal2012}. The key improvement over standard PT consisted in reorganizing the perturbative series in a more efficient way by means of a new building block, the nonlinear propagator, a quantity measuring the response of the density and velocity perturbations to their initial conditions. 
The possibility to analytically compute this new object revolutionized indeed the cosmological perturbation theory approach. 

A first extension of standard PT to the specific case of clustering quintessence can be found in \cite{SefusattiVernizzi2011}, where the pressure-less perfect fluid equations have been worked out, while in \cite{DAmicoSefusatti2011, AnselmiBallesterosPietroni2011} it was shown how to compute the nonlinear power spectrum (PS) by means of the Time-Renormalization Group (TRG) method of \cite{DAmicoSefusatti2011, AnselmiBallesterosPietroni2011}. However, the computation of the nonlinear propagator for these cosmologies was still lacking. This is the first achievement of the present work, allowing the extension of a large set of powerful resummed techniques to clustering Dark Energy models.

We compute the nonlinear PS exploiting the resummation scheme proposed by Anselmi \& Pietroni \cite{AnselmiPietroni2012} (hereafter AP). Such scheme takes into account the relevant diagrams in the perturbative expansion and proposes a new interpolation procedure between small and large scales. What makes AP so appealing is the possibility to extend the predictions, for the first time, to scales beyond the Baryonic Acoustic Oscillation (BAO) range, together with a fast and easy code implementation. It is important to stress that the AP approach, as other perturbative techniques, is intrinsic limited by the emergence of multi-streaming effects  at small scales, i.e. the generation of velocity dispersion due to orbit crossing. Clearly, the breaking-down of the pressureless perfect fluid (PPF) approximation employed as a starting point for the fluid equations has nothing to do with the perturbative scheme used. The AP prediction agrees with N-body simulations at the 2\% level accuracy on BAO scales while at 
smaller 
scales shows discrepancies only of a few percents up to $k \simeq 1\kMpc$ at $z \gtrsim 0.5$. As we expect, the level of the agreement is consistent with the findings of \cite{PueblasScoccimarro2009} where the authors estimated the departures from the PPF assumption by means of high resolution N-body simulations. Clearly, it would be desirable to extend our results beyond the PPF description. However, so far, even in the standard scenario this is still subject of intense investigations (see \cite{PueblasScoccimarro2009, ValageasNishimichi2011A, PietroniEtal2011, CarrascoHertzbergSenatore2012, PajerZaldarriaga2013, ValageasNishimichiTaruya2013} for an incomplete list of contributions). 

Taking advantage of the AP approach, we will show that additional nonlinear corrections due DE clustering become larger than $1\%$ for scales smaller than BAO for 10\% variation from $w=-1$. Thus it allows us to finally test the approximation employed in the TRG predictions of \cite{AnselmiBallesterosPietroni2011} and the assumptions made in several works forecasting the detectability of dark energy models in presence of quintessence perturbations \cite{SaponeKunz2009, AmendolaKunzSapone2008, SaponeKunzAmendola2010}. Furthermore, as a by-product, we provide a theoretically motivated mapping from the nonlinear matter power spectrum in {\em smooth} quintessence models to the nonlinear, total density power spectrum in {\em clustering} quintessence models that could possibly serve as a consistency check for numerical results once these will become available. This is not of small significance, given the challenge posed by accurate simulations of 
structure formation in this kind of models.  
  
This paper is organized as follows. In Section \ref{EqMotionLinear} we review the equations of motion for the quintessence--dark matter fluid. In Section \ref{dynamics} we recast the starting fluid equations in a compact form and make explicit the nonlinear behavior of perturbations. In Section \ref{StatisticalObs}, in order to compute the nonlinear density power spectrum, we apply the selected resummation scheme to the clustering quintessence fluid. In Section \ref{results} we present our results and a well motivated mapping from the smooth to the clustering quintessence PS. Section \ref{discussion} is devoted to our conclusions and possible future developments. Some details about the resummation scheme and a comparison with recent numerical simulations for a $\Lambda$CDM cosmology are included as Appendices.

%%%%%%%%%%%%%%%%%%%%%%%%%%%%%%%%%%%%%%%%%%%%%%%%%%%%%%%%%%%%%%%%%%%%%%%%%%%%%%%%%%%%%%%%%%%%%%%%%
%%%%%%%%%%%%%%%%%%%%%%%%%%%%%%%%%%%%%%%%%%%%%%%%%%%%%%%%%%%%%%%%%%%%%%%%%%%%%%%%%%%%%%%%%%%%%%%%%
\section{Equations of motion and linear solutions}
\label{EqMotionLinear}

The equations of motion of matter and dark energy perturbations for a quintessence field characterized by a vanishing speed of sound, along with their linear solutions, have been studied in several works \cite{SefusattiVernizzi2011, DAmicoSefusatti2011,AnselmiBallesterosPietroni2011}.  In order to keep this work as much self-contained as possible and to set the notation, in this section we will summarize these results and the relevant aspects of the solutions at the linear level.

We consider a spatially flat FRW space-time with metric $ds^2=-a^2(\tau)[-d\tau^2+\delta_{ij}dx^idx^j]$ (where $\tau$ is the conformal time given by $d\tau=d t/a(t)$ and $t$ the cosmic time), containing only cold dark matter (CDM) and dark energy. For sake of simplicity, we assume a DE equation of state with constant  $w= \bar p_Q/\bar \rho_Q$, so that the Hubble  rate is given by
\be
H^2=\lp\frac{1}{a}\frac{da}{dt}\rp^2=H_0^2\lb\Omega_{m,0} a^{-3}+\Omega_{Q,0}a^{-3(1+w)}\rb,
\ee  
where $\Omega_{m,0}=\bar \rho_{m,0}/(\bar \rho_{m,0}+\bar \rho_{Q,0})=1-\Omega_{Q,0}$, with the subscript ``0'' denoting quantities evaluated today, and $\bar \rho_{m}$ ($\bar \rho_{Q})$  being the mean value of the matter (quintessence) energy density $\rho_m$ ($\rho_{Q}$).  

At the level of perturbations, we consider the perfect fluid approximation in the Newtonian regime to describe both the CDM and DE components, assuming the system to interact only gravitationally. In addition, we restrict ourself to models where the rest frame sound speed vanishes, $c_s=0$ (for an analysis where $c_s$ is allowed to take values  between  $c_s=0$ and $c_s=1$ see \cite{AnselmiBallesterosPietroni2011}). In that case the two fluids are comoving and, following the notation of \cite{SefusattiVernizzi2011}, the fluid equations will be given by the Euler equation for the common CDM and quintessence velocity field $\vec v$,
\be
\frac{\partial {\vec{ v}}}{\partial \tau}  + \HH \vec v+ (\vec v \cdot \vec \nabla) \vec v =-\vec \nabla \Phi\,, \label{euler_common}
\ee
and by the continuity equations for the density contrast of dark matter $\delta_m\equiv\delta\rho_m/\bar \rho_m$
and  quintessence $\delta_Q\equiv\delta\rho_Q/\bar \rho_Q$, 
\bea
&&\frac{\partial \delta_m}{\partial \tau}  + \vec \nabla \cdot \big[ (1+\delta_m) \vec v \big] = 0 \label{continuity_m}\,,\\
&&\frac{\partial \delta_Q}{\partial \tau}  -3 w \HH \delta_Q+ \vec \nabla \cdot \big[ (1+w+\delta_Q) \vec v\big] = 0 \label{continuity_Q}\,,
\eea
with $\HH \equiv d \ln a  /d \tau $  the conformal Hubble rate and $\Phi$ the gravitational potential satisfying the Poisson equation,
\be
\nabla^2 \Phi = \frac{3}{2}\HH^2\Omega_m\lp\delta_m+\delta_Q\frac{\Omega_Q}{\Omega_m}\rp= \frac{3}{2}\HH^2\Omega_m\delta, \label{poisson}
\ee where $\Omega_{m,Q}=\Omega_{m,Q}(\tau)\equiv\bar \rho_{m,Q}(\tau)/[\bar \rho_m(\tau)+\bar \rho_m(\tau)]$. In the last equality we have used the convenient definition for the ``total'' density contrast,
\be \delta\equiv \delta_m+\delta_Q\frac{\Omega_Q}{\Omega_m}\,,\ee
corresponding to the weighted sum of matter and quintessence perturbations, normalized such that $\delta\rightarrow\delta_m$ for $\delta_Q\rightarrow0$. 

In Fourier space, by combining the previous equations and introducing the function
\be\label{cc}
C(\tau)=1+(1+w)\frac{\Omega_Q(\tau)}{\Omega_m(\tau)},\ee
it is straightforward to  see that the evolution equations for the total density contrast $\delta$ and velocity divergence $\theta\equiv \vec{\nabla}\cdot \vec v$ can be recast as
\bea
&&\frac{\partial \delta_{\vec k}}{\partial \tau}  + C  \theta_{\vec k} = -\int d^3q_1 d^2q_2\delta_D(\vec k-\vec q_{12})\alpha(\vec q_1, \vec q_2) \theta_{\vec q_1} \delta_{\vec q_2} \label{continuity_tot}\,,\\
&&\frac{\partial \theta_{\vec k}}{\partial \tau} + \HH \theta_{\vec k} + \frac{3}{2} \Omega_m \HH^2 \delta_{\vec k} = -\int d^3q_1 d^2q_2\delta_D(\vec k-\vec q_{12}) \beta(\vec q_1, \vec q_2) \theta_{\vec q_1} \theta_{\vec q_2}\;, \label{euler_tot}
\eea with
\bea
\alpha(\vec q_1, \vec q_2) & \equiv & 1 + \frac{\vec q_1 \cdot \vec q_2}{q_1^2}  \;,  \label{alpha_def}\\
\beta(\vec q_1, \vec q_2)&\equiv& \frac{(\vec q_1 + \vec q_2)^2 \,\vec q_1 \cdot \vec q_2 }{2\, q_1^2\, q_2^2}\;,\label{beta_def}
\eea
where we adopt the notation $\vec q_{ij}\equiv\vec q_i+\vec q_j$ and $q_i\equiv|\vec q_i|$.

\begin{figure}[t]
\begin{center}
\includegraphics[width=0.49\textwidth]{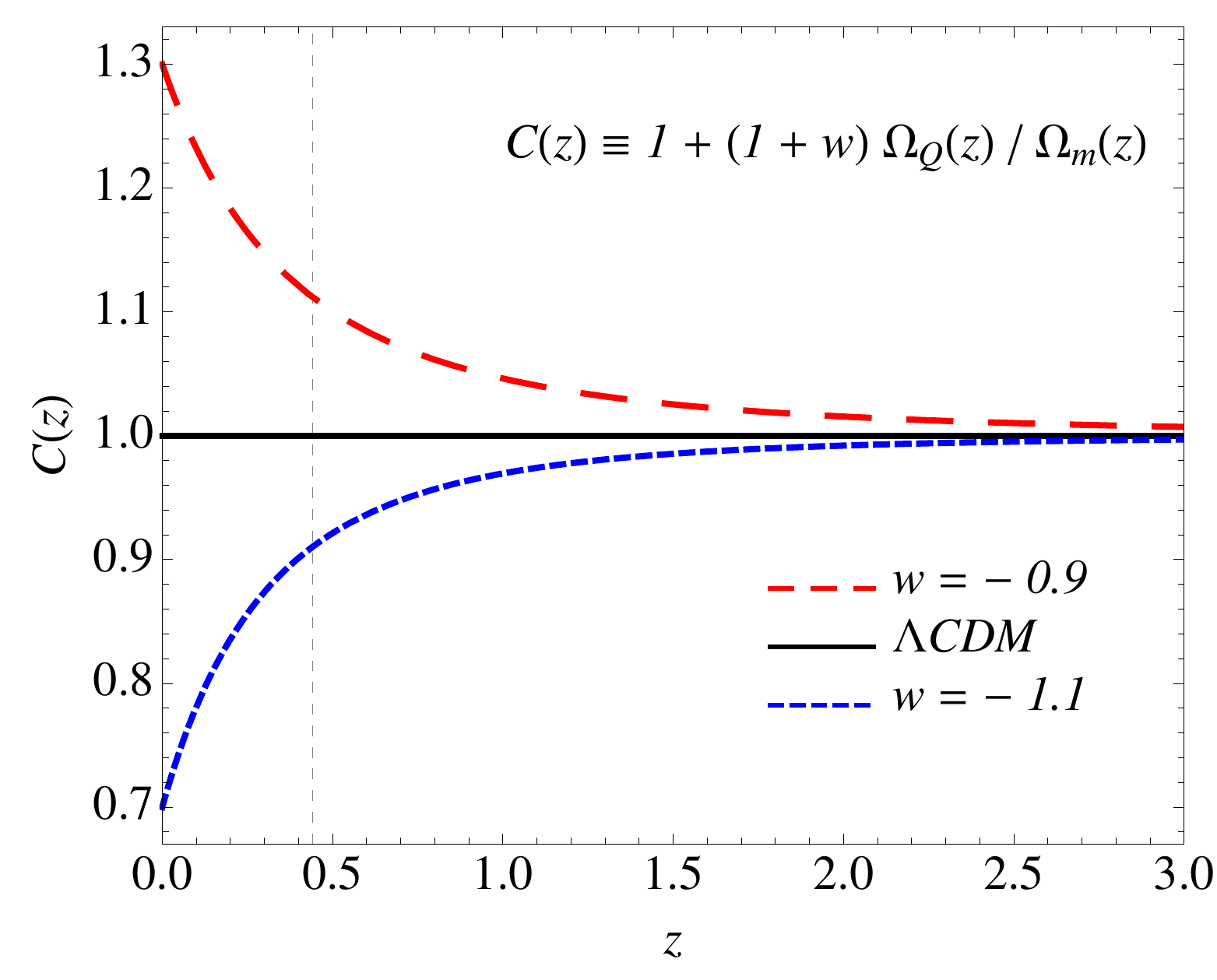}
\caption{The function $C$ defined in eq.~(\ref{cc}) as a function of redshift for $\Omega_{m,0}=0.25$,  $w=-1.1$ ({\em blue, short-dashed line}), $w=-1$ ($\Lambda$CDM,  {\em black, solid line}), and $w=-0.9$ ({\em red, long-dashed line}). The vertical line indicates the redshift of equality between matter and quintessence for the $\Lambda$CDM
cosmology.}
\label{Fig1}
\end{center}
\end{figure}

Note that for $C(\tau)=1$ and $\delta_Q=0$ (i.e., $\delta=\delta_m$) these equations reduce to the usual equations that describe the matter density contrast $\delta_m$ in the smooth quintessence scenario (with $c_s=1$). If, in addition, we set $w=-1$, we reduce to the case of a $\Lambda CDM$ cosmology. In other terms, the function $C(\tau)$ captures all the corrections to the equations of motion induced by the clustering of quintessence. It is therefore crucial, for the interpretations of our results, to keep in mind the behavior of this function for different values of $w$. This is shown in Fig. \ref{Fig1} (reproducing Fig.~1 of \cite{SefusattiVernizzi2011, DAmicoSefusatti2011}), where $C(\tau)$ is plotted as a function of 
redshift for 
$\Omega_{m,0}=0.25$ and different values of $w$.  

\begin{figure}[t]
\begin{center}
\includegraphics[width=0.98\textwidth]{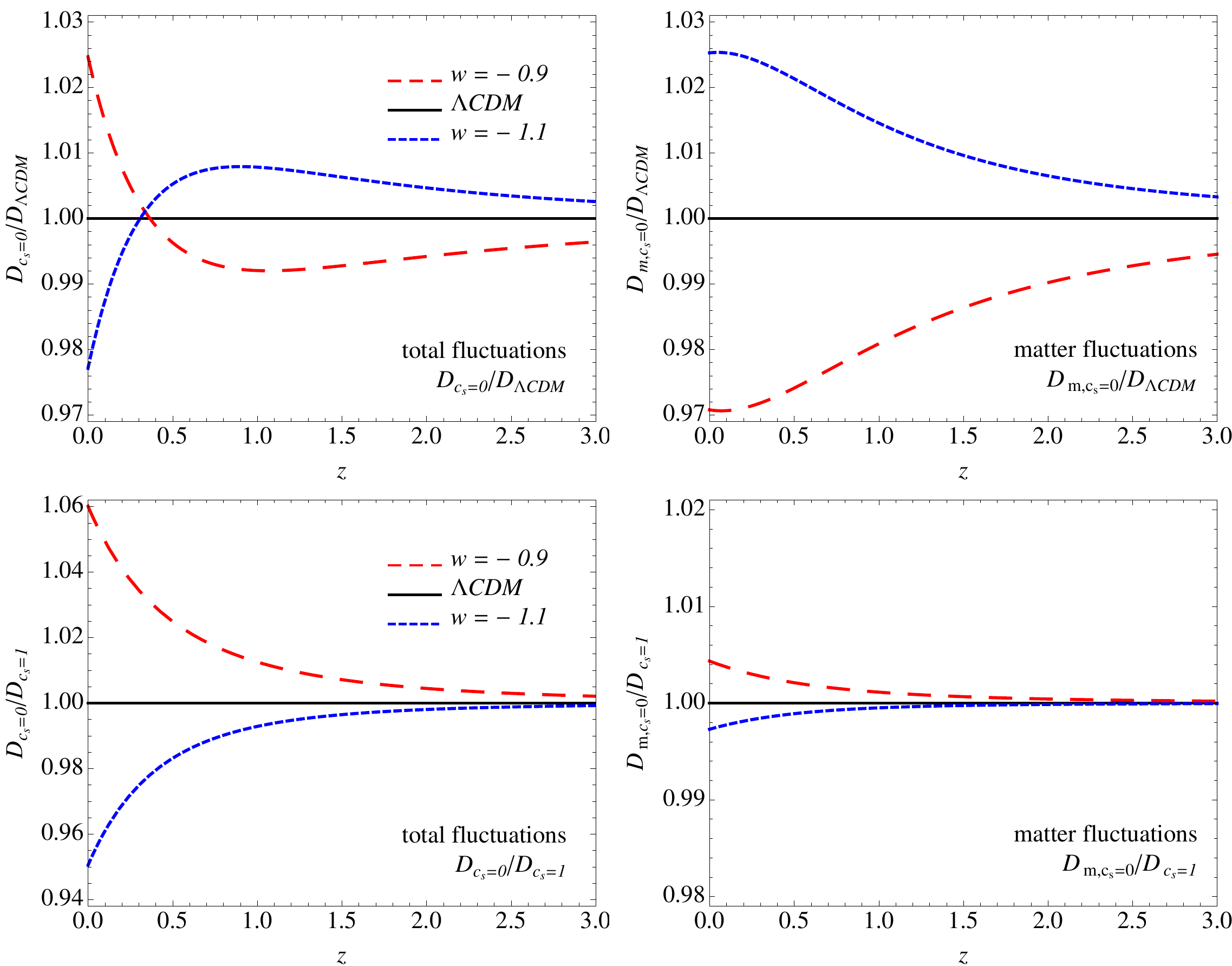}
\caption{Linear growth factor for the matter, $D_{m, c_s=0}$ ({\em right panels}), and total, 
$D_{c_s=0}$  ({\em left panels}), density perturbations as a function of the redshift, for $w=-1.1$ ({\em blue, short-dashed line}) and $w=-0.9$ 
({\em red, long-dashed line}), $w=-1$ ($\Lambda$CDM, {\em black, solid line}), shown as the ratio to the $\Lambda$CDM, $w=-1$ case ({\em top panels}) 
and as the ratio to the growth factor $D_{c_s=1}$ corresponding to the smooth case, $c_s=1$ ({\em bottom panels}). All curves correspond to $\Omega_{m,0}=0.25$.}
\label{FigGrowthFact}
\end{center}
\end{figure}
  
Following \cite{SefusattiVernizzi2011}, the linear solutions for the total density contrast and for the velocity divergence can be written in the separable form
\bea
\delta^{\rm lin}_{\vec k} (\tau) &\equiv &D(\tau) \delta^{\rm in}_{\vec k}\;,\label{delta_lin}\\
\theta^{\rm lin}_{\vec k}(\tau) &\equiv &- \frac{\HH(\tau) f(\tau)}{C(\tau)} D(\tau) \delta^{\rm in}_{\vec k}\;, \label{theta_D}
\eea
where $D$ is the linear growth function and $f$ the linear growth rate, 
\be 
f\equiv\frac{d \ln D}{d \ln a}\,.
\ee

For the smooth case ($c_s=1$) and {\em constant} equation of state parameter $w$, the solution for the linear growing mode is known in terms of hypergeometric functions as \cite{SilveiraWaga1994, Padmanabhan2003}
\be
\frac{D_{c_s=1}(z)}{a}=\,{}_2F_1\left(-\frac1{3w},\frac{-1 + w}{2 w},1 - \frac5{6 w},
-x\right)\,,
\ee
%\bea
%\frac{D_{+,c_s=1}(a)}{a}&=&\frac{ 2^{-\frac{w+1}{w}} x^{\frac{5}{6 w}} (x+1)^{-\frac{w+1}{2 w}} \sec \left(\frac{\pi }{6 w}\right) \Gamma \left(1-\frac{5}{6 w}\right) }{\sqrt{\pi } \Gamma
%   \left(\frac{w-1}{w}\right)}\nonumber\\
%   &\times&\left\{2 \pi 
%   \sqrt{x+1}\,{}_2\tilde{F}_1\left(\frac{1}{2w},\frac{1+w}{2},\frac{1}{2}+\frac{1}{3w};\frac{1}{x+1}\right)\right.\nonumber\\
%   &+& 2^{\frac{1}{3 w}} (x+1)^{\frac{1}{6 w}} \sin \left(\frac{2 \pi }{3 w}\right) \Gamma \left(1+\frac{2}{3 w}\right) \Gamma \left(\frac{w-1}{w}\right) \nonumber \\
%   &\times&\left. \,{}_2\tilde{F}_1\left(\frac{1}{2}+\frac{1}{3w},1+\frac{1}{3w},\frac{3}{2}-\frac{1}{6 w};\frac{1}{x+1}\right)\right\}\,.  
%   \eea
with
\be\label{eq:x}
x\equiv \frac{\Omega_Q(\tau)}{\Omega_m(\tau)}=\frac{(1-\Omega_{m,0})}{\Omega_{m,0}}a^{-3w}\,,
\ee
and where $D$ is normalized such that  $D(a)/a\to 1$ as $a\to 0$, recovering the behavior $D(a)\sim a$ at early time, during matter domination. 

In the case of clustering quintessence with $c_s=0$, for constant $w$, the equations always admit an integral solution for the growing mode $D_{c_s=0}$ of the {\em total} density fluctuations, given by \cite{SefusattiVernizzi2011}
\be
D_{c_s=0}(a) = \frac52\, H_0^2\, \Omega_{m,0}\, H(a)\! \int_0^a\!\! \frac{C(\tilde a)}{[\tilde a\, H(\tilde a)]^3} d \,\tilde a\;. \label{int_sol}
\ee
It can also be expressed in terms of Hypergeometric functions as
\be
\frac{D_{+,c_s=0}(a)}{a}=\frac{5 (3 w-2 x-2)}{9 w}-\frac{2 (3 w-5) \sqrt{x+1}\,\,{}_2F_1\left(-\frac{1}{2},-\frac{5}{6 w},1-\frac{5}{6 w};-x\right)}{9 w},
\ee 
with $x$ defined as in eq.~(\ref{eq:x}) above.

The linear growth factor for the {\em matter} perturbations alone $D_{m,c_s=0}$ can be obtained from the solution for the total density, given the relation  \cite{SefusattiVernizzi2011}:
\be
\frac{dD_{m,c_s=0}}{d\ln a\,\,}=\frac{1}{C}\frac{dD_{c_s=0}}{d\ln a}\,.\label{dmplus}
\ee

These different solutions are shown in Fig.~\ref{FigGrowthFact} as a function of redshift, for $w=-1.1$ and $w=-0.9$. Top panels plot the ratio between the growth factor for the clustering case $c_s=0$ to the one for the $\Lambda$CDM model, while bottom panels plot the ratio of the growth function for the clustering case to the corresponding solution for smooth quintessence\footnote{Note that while there are theoretical motivations for considering $w<-1$  for $c_s\to 0$,  this is not the case for $c_s=1$, due to the presence of ghost instabilities. However, we consider also the latter case just for comparison.}. Left and right panels show, respectively, the results for the total and matter perturbations.

The upper left panel, in particular, illustrates how different and competing effects are at play in the clustering quintessence models. Since the fraction $\Omega_Q/\Omega_m$  increases as $(z+1)^{-3|w|}$ when $z$ decreases, if $w>-1$ ($w<-1$ leads to the opposite results) the redshift at which the acceleration becomes relevant (i.e. when $\Omega_Q/\Omega_m \simeq 1$) is larger than for $\Lambda$CDM, suppressing more the growth of density perturbations.  However, in the case of clustering quintessence with $w>-1$ ($w<-1$ ), the presence of quintessence fluctuations leads to a larger (smaller) growth for the total perturbations. This second effect dominates at low redshift for the total density perturbations, while it is quite subdominant  for the matter perturbations. In fact, $D_{c_s=0}/D_{c_s=1} \sim 5\%$ while $D_{m,c_s=0} \sim D_{c_s=1}$ at better that $1\%$ at $z=0$ for $w=-1.1$ or $-0.9$.

%%%%%%%%%%%%%%%%%%%%%%%%%%%%%%%%%%%%%%%%%%%%%%%%%%%%%%%%%%%%%%%%%%%%%%%%%%%%%%%%%%%%%%%%%%%%%%%%%
%%%%%%%%%%%%%%%%%%%%%%%%%%%%%%%%%%%%%%%%%%%%%%%%%%%%%%%%%%%%%%%%%%%%%%%%%%%%%%%%%%%%%%%%%%%%%%%%%
\section{Nonlinear solutions: an analytical insight}
\label{dynamics}

In this section we show how it is possible to obtain some insights on the nonlinear evolution of the total density field, simply by inspection of the formal solution to the equations of motion. Following the notation of \cite{Scoccimarro1997}, we rewrite the fluid equations using\footnote{From now on, unless explicitly stated, we will refer to $D$ and $D_m$ as the linear growth factors for total and matter perturbations respectively in the clustering DE scenario, i.e., we drop the ``$c_s=0$'' subscript.} $\eta=\log [ D(z)/D(z_{in})]$ as the time variable, and introduce the doublet $\varphi_a$ ($a=1,2$), defined by
\be\left(\begin{array}{c}
\varphi_1 ( {\vec k}, \eta)\\
\varphi_2 ( {\vec k}, \eta)  
\end{array}\right)
\equiv 
e^{-\eta} \left( \begin{array}{c}
\delta_{\vec k}  ( \eta) \\
-\frac{C}{\HH f}\theta_{\vec k}(\eta)
\end{array}
\right)\,,
\label{doppietto}
\ee along with the {\it vertex} matrix, $\gamma_{abc}({\vec k},{\vec p},{\vec q}) $ ($a,b,c,=1,2$) 
whose only independent, non-vanishing, elements are
\bea
&&\gamma_{121}({\vec k},\,{\vec p},\,{\vec q}) = 
\frac{1}{2} \,\delta_D ({\vec k}+{\vec p}+{\vec q})\, 
\alpha(\vec p,\vec q)\,,\nonumber\\
&&\gamma_{222}({\vec k},\,{\vec p},\,{\vec q}) = 
\delta_D ({\vec k}+{\vec p}+{\vec q})\, \beta(\vec p,\vec q)\,,
\label{vertice}
\eea
and 
$\gamma_{121}({\vec k},\,{\vec p},\,{\vec q})  = 
\gamma_{112}({\vec k},\,{\vec q},\,{\vec p}) $.
Then, the two equations (\ref{continuity_tot}) and (\ref{euler_tot}) can  be recast as
\be
\partial_\eta\,\varphi_a({\vec k}, \eta)= -\Omega_{ab}\,
\varphi_b({\vec k}, \eta) +\frac{ e^\eta}{C(\eta)} 
\gamma_{abc}({\vec k},\,-{\vec p},\,-{\vec q})  
\varphi_b({\vec p}, \eta )\,\varphi_c({\vec q}, \eta )\,,
\label{compact}
\ee
with
\be
\Omega= \left(\begin{array}{cc}
\ds 1 & \ds -1\\&\\
\ds -\frac{3\,\Omega_m\, C}{2f^2} & \ds \frac{3\,\Omega_m\, C}{2f^2} \end{array}
\right)\,,
\label{bigomega}
\ee 
and where repeated indexes are summed over and integration over momenta $\vec q$ and $\vec{p}$ is implied. Again, for $C=1$ and $\delta=\delta_m$ eq.~(\ref{compact}), with eq.~(\ref{bigomega}), reduces to the corresponding one for the case of smooth DE, after replacing the functions $\eta$, $\HH$ and $f$ in eq.~(\ref{doppietto}) by the appropriate ones.

As for $\Lambda$CDM cosmologies, what complicates the implementation of a resummation technique to these equations is
the fact that the $\Omega$ matrix is time dependent. However,  as for the $\Lambda$CDM case \cite{CrocceScoccimarro2006B},
most of the time  we have $\Omega_m C/f^2\simeq 1$, and one can see that for $w>-1$ the approximation $\Omega_m C/f^2\simeq 1$ works  better than for $\Lambda$CDM, while it worsen for $w<-1$. In other words, this means that at linear level almost all of the information about the cosmological parameters is encoded in the linear growth factor $D$ which has been rescaled out in eq.~(\ref{doppietto}).
Therefore, as commonly done for  $\Lambda$CDM cosmologies, in what follows  we consider the approximation
\be
\Omega\simeq \left(\begin{array}{cc}
\ds 1 & \ds -1\\&\\
\ds -\frac{3 }{2} & \ds \frac{3}{2} \end{array}
\right)\,,
\label{bigomegaap} 
\ee 
which is expected to be accurate at better than $1\%$ for $k<0.2\kMpc$, but quickly improving at larger redshift (see, in particular, Appendix A in \cite{CrocceScoccimarroBernardeau2012} for a relevant test).

The linear solution $\varphi^{L}$ can be expressed as
\be
\varphi^{L}_{a}(\vec k,\eta)=g_{ab}(\eta,\eta_{in})\varphi^{L}_{b}(\vec k,\eta_{in})\,,
\ee where  $g_{ab}$ is the retarded  linear propagator which  obeys the equation 
\be\label{eqlinprop}
(\delta_{ab}\partial_{\eta}+\Omega_{ab})g_{bc}(\eta,\eta_{in})=\delta_{ac}\delta_D(\eta-\eta_{in})\,,
\ee
with causal boundary conditions. Explicitly, it is given  by  \cite{CrocceScoccimarro2006B}:
\be
g_{ab}(\eta,\eta^\prime) =\left[ {\bf B} + {\bf A}\, e^{-5/2 
(\eta -\eta^\prime)}\right]_{ab}\, \theta(\eta-\eta^\prime)\,,
\label{proplin}
\ee
 with $\theta$ the 
step-function, and
\be {\bf B} = \frac{1}{5}\left(\begin{array}{cc}
3 & 2\\
3 & 2
\end{array}\right)\,\qquad {\mathrm{and}} \qquad
{\bf A} = \frac{1}{5}\left(\begin{array}{cc}
2 & -2\\
-3 & 3
\end{array}\right)\,.\ee
The initial conditions corresponding respectively to the growing and decaying modes can be selected by 
considering initial fields $\varphi_a$ proportional to 
\be u_a = \left(\begin{array}{c} 1\\ 
1\end{array}\right)\,\qquad\mathrm{and} 
\qquad v_a=\left(\begin{array}{c} 1\\ -3/2\end{array}\right)\,.
\label{ic}
\ee 

At this stage, before introducing statistics and computing the observables, we can extract the first insights about the non--linear behavior of the clustering quintessence fluid. The formal solution of eq.~(\ref{compact}) reads
\be
\varphi_a({\vec k}, \eta_a)=g_{ab}(\eta_a,\eta_b) \varphi^{0}_b({\vec k}, \eta_b)+\int_{\eta_b}^{\eta_a} d s\,\frac{ e^s}{C(s)} g_{ab}(\eta_a,s) \gamma_{bcd}({\vec k},\,-{\vec p},\,-{\vec q}) \varphi_c({\vec p}, s) \varphi_d({\vec q}, s)\,,
\label{formal}
\ee
where, assuming growing mode initial conditions and being $\delta_{0}$ the initial value of the field, we have $g_{ab}(\eta_a,\eta_b) \varphi^{0}_b({\vec k}, \eta_b)=\delta_{0}({\vec k}) u_b$ . 
Following \cite{CrocceScoccimarro2006A} we can expand $\varphi_a({\vec k}, \eta_a)$ in this fashion
\be
\varphi_a({\vec k}, \eta_a)=\sum_{n=1}^{\infty}\varphi_a^{(n)}({\vec k},\eta_a)
\label{phin}
\ee
with
\be
	\vp_{a}^{(n)}({\vec k},\eta_{a})=\int d^{3}q_{1}\cdots d^{3}q_{n}\delta_{D}({\vec k}-{\vec q}_{1\dots n})\mathcal{F}_{a}^{(n)}({\vec q}_{1},\dots,{\vec q}_{n};\eta_{a})\delta_{0}({\vec q}_{1})\cdots\delta_{0}({\vec q}_{n})\, 
	\label{vpn}
\ee
where ${\vec q}_{1\dots n}\equiv {\vec q}_{1}+\cdots+{\vec q}_{n}$. Replacing (\ref{phin}) and (\ref{vpn}) in (\ref{formal}) we find the kernel recursion relations 
\bea
	  &&\mathcal{F}^{(n)}_{a}({\vec q}_{1},\dots,{\vec q}_{n};\eta_{a})\delta_{D}({\vec q}-{\vec q}_{1\dots n})=\left[\sum_{m=1}^{n}\int_{\eta_b}^{\eta_{a}} d s \;\frac{e^{s}}{C(s)}\; g_{ab}(\eta_a,s)\gamma_{bcd}({\vec k},-{\vec q}_{1\dots m},-{\vec q}_{m+1\dots n})\right.  \nonumber\\
	  &&\qquad\qquad\left.\times\mathcal{F}_{c}^{(m)}({\vec q}_{1},\dots,{\vec q}_{m}; s)\mathcal{F}_{d}^{(n-m)}({\vec q}_{m+1},\dots,{\vec q}_{n}; s) \right]_{\textrm{sym}}\, ,
	  \label{PTrecursion}
\eea
where the rhs has to be symmetrized under interchange of any two wave vectors. For $n=1$, $\mathcal{F}^{(1)}(\eta_{a})=g(\eta_{a},\eta_{b})u_{b}$.  For a
$\Lambda$CDM cosmology, in the limit where the initial conditions are imposed in the infinite past $\eta_{b}\rightarrow-\infty$ we  recover the well known SPT recursion relation \cite{CrocceScoccimarro2006A}. On the other hand in the clustering quintessence case we can not analytically perform the time integrals. However, when we take into account just the propagator growing mode in eq. (\ref{PTrecursion}), the times integrals will contribute a factor \footnote{The other terms contributing to $\mathcal{F}^{(n)}_{a}$, given by the propagator decaying mode, will be of the same order.}
\be
\int_{\eta_b}^{\eta_a}ds_1\int_{\eta_b}^{s_1}ds_2 ...\int_{\eta_b}^{s_{n-1}} ds_{n} \prod_{j=1}^{n}\frac{e^{s_{j}}}{C(s_{j})} \, .
 \ee
Recalling that from eq. (\ref{dmplus}) we have
\be
\int_{\eta_b}^{\eta_a}ds\, \frac{e^{s}}{C(s)}=\frac{D_{m}(\eta_a)-D_{m}(\eta_b)}{D_{in}}\, ,
%\label{Idef}
 \ee
where $D_{in}=D(\eta_{in})$ is the growth factor at some initial time $\eta_{in}$. It follows that
\be
	\mathcal{F}^{(n)}_{a}({\vec k}_{1},\dots,\bk_{n};\eta_{a})\,\delta_{D}({\vec k}-{\vec k}_{1\dots n})\sim D_{m}^{(n-1)}(\eta_a).
	\label{scalingFn}
 \ee
 Therefore the $n$-order contribution of the total density contrast follows $\delta_{\vec k}^{(n)} \sim D\, D_{m}^{(n-1)}$ as was already pointed out at second order in \cite{DAmicoSefusatti2011}. Basically this means that the non-linear interactions are driven by dark matter,  while the effect of DE enters as a multiplicative factor (i.e. $D$), acting in the same way at both linear and non--linear level. The resummation we will perform confirms this behavior.

%%%%%%%%%%%%%%%%%%%%%%%%%%%%%%%%%%%%%%%%%%%%%%%%%%%%%%%%%%%%%%%%%%%%%%%%%%%%%%%%%%%%%%%%%%%%%%%%%
%%%%%%%%%%%%%%%%%%%%%%%%%%%%%%%%%%%%%%%%%%%%%%%%%%%%%%%%%%%%%%%%%%%%%%%%%%%%%%%%%%%%%%%%%%%%%%%%%
\section{The power spectrum in the mildly nonlinear regime}
\label{StatisticalObs}

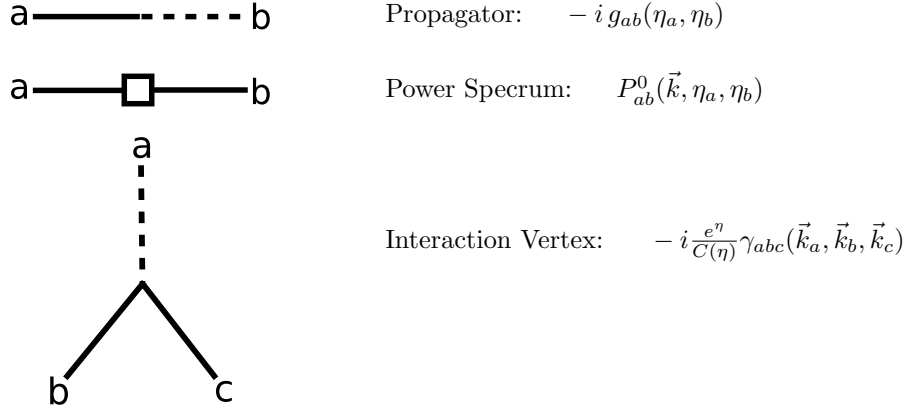
\begin{figure}[t]
\vskip 0.2cm
\begin{center}
\begin{flushleft}
 \setlength{\unitlength}{1cm}
\begin{picture}(0,0)
\put(1.5,-0.2){\includegraphics[width=3.5cm]{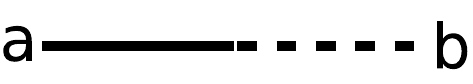}}
\put(6.5,-0){$\mbox{Propagator:}\,\,\,\,\,\,\,\,\,-i\,g_{ab}(\eta_a,\eta_b)$}
\put(1.5,-1.2){\includegraphics[width=3.5cm]{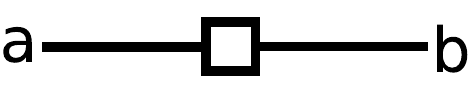}}
\put(6.5,-1){$\mbox{Power Specrum:}\,\,\,\,\,\,\,\,\,\,P_{ab}^0(\vec{k},\eta_a,\eta_b)$}
\put(2,-5.2){\includegraphics[width=2.5cm]{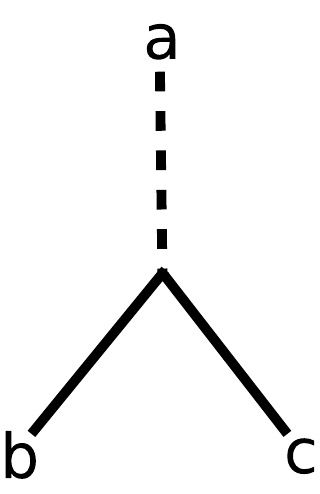}}
\put(6.5,-3){$\mbox{Interaction Vertex:}\,\,\,\,\,\,\,\,\,\,-i \frac{e^\eta}{C(\eta)} \gamma_{abc}(\vec{k}_a,\vec{k}_b,\vec{k}_c)$}
\end{picture}
\vskip 5cm
\end{flushleft}
\end{center}
\caption{ Feynman rules for cosmological perturbation theory with clustering quintessence}
\label{FeynmanRules}\end{figure}

We focus now on the nonlinear behavior of a fundamental quantity: the total density power spectrum.
In this section  we present the nonlinear evolution equations which we then solve numerically to compute the power spectrum. 
To this end we use of the resummation scheme proposed in \cite{AnselmiPietroni2012, AnselmiMatarresePietroni2011}.  
Although a detailed description of the resummation scheme can be found in these references, for the sake of 
completeness, in Appendices \ref{appLargek} and \ref{EvolutEq} we include a summary of the main relevant aspects of it applied to the clustering quintessence case. 
The generalization of this approach 
to the clustering quintessence case is quite straightforward because, in the absence of a sound horizon, there is no additional preferred scale 
in the system, and moreover, as described in the previous section, in the perfect fluid approximation the velocity field is the same for
both DM and DE, allowing us to reduce the number
of equations to those of a single fluid. Indeed, given the linear propagator approximation introduced in the previous section, the only
difference in the implementation of the diagrammatic technique developed for the $\Lambda$CDM  case in \cite{CrocceScoccimarro2006B}, and
of the functional method of \cite{MatarresePietroni2007}, is that now each vertex comes multiplied by the time dependent 
function $C(\eta)$, as $e^\eta\,\rightarrow\,e^\eta\,/\,C(\eta)$. 
In what follows,  whenever we make use of the diagrammatic representations for specific perturbative contributions, we refer to the definition of the Feynman rules in Fig.  \ref{FeynmanRules}, where $P^0_{ab}$ is the linear power spectrum,
\be
P^0_{ab}(\vec{k},\eta_a,\eta_b)=g_{ac}(\eta_a,\eta_{in})g_{bd}(\eta_b,\eta_{in})P^0_{cd}(\vec{k},\eta_{in},\eta_{in}),
\ee
 which we define  at the initial time $\eta_{in}$ in terms of the linear growing mode as $P_{ab}^{0}(\vec{k},\eta_{in},\eta_{in})\simeq P_{in}(k)u_a u_b$, with $ P_{in}(k)$ the initial density power spectrum.

\begin{figure}[t]
\begin{center}
\includegraphics[width=0.7\textwidth]{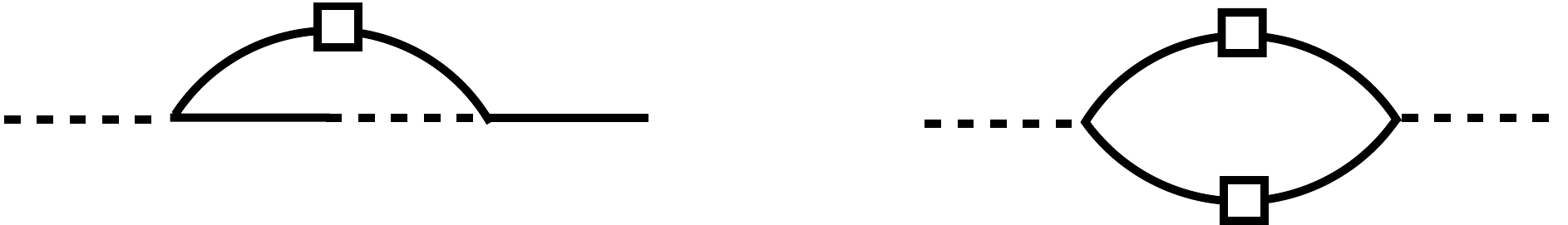}
\end{center}
\caption{ Left: $\Sigma_{ab}^{(1)}$, the 1-loop contribution to  $\Sigma_{ab}$. Right: $\Phi_{ab}^{(1)}$, the 1-loop contribution to $\Phi_{ab}$.}
\label{SigmaPhi}
\end{figure}

At nonlinear level, it is well-known that either by exploring the diagrammatic structure of the contributions at arbitrary high order \cite{CrocceScoccimarro2006A} or by applying  functional methods \cite{MatarresePietroni2007}, it is possible to express the exact evolution equations for the propagator and for the PS, as
\bea
 &&G_{ab} (k;\, \eta_a, \eta_b)=g_{ab} (\eta_a-\eta_b) \nonumber\\
 &&\qquad\qquad+ \int_{\eta_b}^{\eta_a} d s\,\int_{\eta_b}^s
d s^\prime\; g_{ac} (\eta_a-s) \Sigma_{cd}(k;\, s, s^\prime) G_{db} (s^\prime - \eta_b),
\label{GEXAT}
\eea
and
\bea
P_{ab}(k; \eta_a,\eta_b) &=& G_{ac}(k;\eta_a,\eta_{in})
G_{bd}(k;\eta_b,\eta_{in}) P_{cd}(k; \eta_{in},\eta_{in})\,\nonumber\\
&&+
 \int_{\eta_{in}}^{\eta_a} d s\,\int_{\eta_{in}}^{\eta_b}
d s^\prime\;
G_{ac}(k;\eta_a,s)
G_{bd}(k;\eta_b, s^\prime) 
\Phi_{cd}(k; s, s^\prime)\,.
\label{fullP}
\eea 
Here the quantity $\Sigma_{cd}(k;\, s, s^\prime)$, known as the ``self energy'', is a causal kernel  ($s^\prime<s$)  given by the  sum of one-particle-irreducible (1PI) diagrams  \footnote{ A 1PI diagram is one that cannot be separated into two disjoint parts by cutting one of its lines.} connecting a   dashed line at time $s$ with a continuous one at $s^\prime$.  The kernel $\Phi_{cd}(k; s, s^\prime)$ is given by the the sum of 1PI diagrams that connect  two dashed lines at $s$ and $s^\prime$ with no time ordering.  The corresponding 1-loop diagrams are shown in Fig. \ref{SigmaPhi}.

In particular, the starting point of \cite{AnselmiPietroni2012,AnselmiMatarresePietroni2011} is the equation for the propagator obtained by deriving eq.~ (\ref{GEXAT}) with respect to $\eta_a$ 
\be
\partial_{\eta_a}\, G_{ab} (k;\, \eta_a, \eta_b) = \delta_{ab}\, \delta_D(\eta_a-\eta_b) - \Omega_{ac} \,G_{cb} (k;\, \eta_a, \eta_b)+\Delta G_{ab}(k;\,\eta_a,\eta_b)\,,
\label{dgtot}
\ee
where
\be
\Delta G_{ab}(k;\,\eta_a,\eta_b)\equiv \int ds^\prime \; \Sigma_{ad}(k;\, \eta_a, s^\prime)\,G_{db} (k;\, s^\prime, \eta_b)\,,
\label{deltaG}
\ee
and the equation obtained deriving eq.~(\ref{fullP}) with respect to $\eta_a$ {\em at equal times}, i.e. after setting $\eta_b=\eta_a$,
\bea
&&\partial_{\eta_a} \,P_{ab}(k; \eta_a) = -\Omega_{ac} \,P_{cb}(k; \eta_a)  -\Omega_{bc} \,P_{ac}(k; \eta_a)\nonumber \\
&&\;\; +\Big[\Delta G_{ac}(k;\eta_a,\eta_{in})  G_{bd }(k;\eta_a,\eta_{in})+  G_{ac}(k;\eta_a,\eta_{in})  \Delta G_{bd }(k;\eta_a,\eta_{in})\Big] P_{in}(k) u_c u_d 
\nonumber\\
&&\;\;+ \int \, ds'\;\big[\Phi_{ac}(k;\eta_a,s') G_{bc}(k;\,\eta_a,s')+ G_{ac}(k;\,\eta_a,s')\Phi_{cb}(k;s',\eta_a) \big]\,\nonumber\\
&&+ \int \, ds\, ds' \;\Phi_{cd}(k;s,s') \Big[\Delta G_{ac}(k;\eta_a,s) G_{bd}(k;\,\eta_a,s')
%\nonumber\\&&\qquad\qquad\qquad\qquad
+  G_{ac}(k;\eta_a,s) \Delta G_{bd}(k;\,\eta_a,s') \Big]\,.
\label{Texact}
\eea  

These are non-local integro-differential equations which are very difficult to solve, unless some approximation is considered. The approximation considered in   \cite{AnselmiMatarresePietroni2011}
relies on the observation that the equations simplify considerably in the limit of large and small values of $k$, presenting a common structure in both limits. Indeed, it has been 
shown that in the large $k$ (or eikonal) limit, the kernel $\Delta G_{ac}$  factorize as (see details in appendices \ref{appLargek} and \ref{appFact})
\be
\Delta G_{ac}(k;\,\eta,s) \simeq H_{{\bf a}}(k;\, \eta,s)\, G_{{\bf a}c}(k;\,\eta,s)\, ,
\label{TRGL}
\ee
where the bold indexes are not summed over, and  
\be\label{Ha} H_{{ a}}(k;\, \eta,s) \equiv  \int_{s}^{\eta} d s''\, \Sigma_{{ a}e}^{(1)}( k;\,\eta\,,s'')\,u_e\,,\ee
with $\Sigma_{{ a}d}^{(1)}$ being the 1-loop approximation to the full $\Sigma_{ad}$, given by (see also Fig.~\ref{SigmaPhi})
\bea
&&\Sigma_{ab}^{(1)}(k;\eta_a,\eta_b) =\nonumber\\
&&\quad  4 \frac{e^{\eta_a+\eta_b}}{C(\eta_a)C(\eta_b)} \int d^3 q \gamma_{acd}(\vec{k},-\vec{q},\vec{q}-\vec{k}) u_c P_{in}(q) u_e \gamma_{feb}(\vec{k}-\vec{q},\vec{q},-\vec{k}) g_{df}(\eta_a,\eta_b)\,.
\label{sigform}
\eea
Ref.~\cite{AnselmiMatarresePietroni2011} also shows that in the opposite limit $k\to 0$, where linear theory is recovered, the
same factorization applies to a very good approximation\footnote{More specifically,  Ref.~\cite{AnselmiMatarresePietroni2011}   shows that the factorization occurs for small values of $k$
but after contracting with one $u_c$,
$\Delta G_{ac}(k;\,\eta,s) u_c\simeq H_{{\bf a}}(k;\, \eta,s)\, G_{{\bf a}c}(k;\,\eta,s)u_c$.
However, it argues that without contracting with $u_b$ the factorization works for the individual components of the propagators in the limit $s\to +\infty$,
and that for $s$ finite  it is still a very good approximation. }.
Therefore, a natural way of interpolating between the small $k$ and large $k$ regimes consists 
in assuming eq.~(\ref{TRGL}) to hold also for intermediate values of $k$, leading to a local equation for the propagator
\be
\partial_{\eta_a}\, \bar{G}_{ab} (k;\, \eta_a, \eta_b) = \delta_{ab}\, \delta_D(\eta_a-\eta_b) - \Omega_{ac} \, \bar{G}_{cb} (k;\, \eta_a, \eta_b)+H_{{\bf a}}(k;\, \eta,s)\,  \bar{G}_{{\bf a}b}(k;\,\eta,s)\,.
\label{dbargtot}
\ee
In fact, the solution of this equation (which we denote as $\bar G$) is exact in the large $k$ limit, and reduces to the 1-loop propagator for  $k\to 0$. 
Computing this solution in our case we obtain, in the large $k$ limit (see Appendix \ref{appLargek}) 
\be
\bar G_{ab}(k, \eta_a, \eta_b)\to    \,G_{ab}^{eik}(k,\eta_a,\eta_b)=g_{ab}(\eta_a,\eta_b)\exp\left[-\frac{1}{2}k^2 \sigma_v^2{\cal{I}}^2(\eta_a,\eta_b)\right].\label{Glargek1}
\ee with the superscript ``eik'' standing for the eikonal limit, 
\be
{ \cal{I}}(\eta_a,\eta_b)=\int_{\eta_b}^{\eta_a}ds\, \frac{e^{s}}{C(s)}=\frac{D_{m}(\eta_a)-D_{m}(\eta_b)}{D_{in}}
 \ee 
where $D_{m}(\eta)$ is defined by eq. (\ref{dmplus}), and
\be  
\sigma_v^2\equiv\frac{1}{3}\int d^3q \frac{P_{in}(q)}{q^2}. \label{sigmav}
\ee  
Therefore, we recover the well-known Gaussian decay found in \cite{CrocceScoccimarro2006B} for the matter propagator in the standard scenario, which takes into account non-perturbatively the main infrared effects of the long-wavelength modes. In the clustering quintessence case, a crucial difference is given by the fact that the derived propagator for the total density field, depends on the matter linear growth factor $D_m$. This is exactly what we expected from the analysis of the previous section.
  
%On the other hand,  for $k\to 0$ we have
%\be
%\bar G_{ab}(k, \eta_a, \eta_b)\to g_{ab}(\eta_a,\eta_b)+\int ds\int ds' g_{ac}(\eta_a,s)\Sigma_{cd}^{(1)}(k,s,s')g_{db}(s',\eta_b), \label{Gsmallk}
%\ee where $\Sigma_{cd}^{(1)}$ is   given in  eq. (\ref{sigform}).

In  \cite{AnselmiPietroni2012}, a similar procedure was followed to obtain an approximation for the evolution equation of the power spectrum. 
Of course, in the physical situation we are considering, the large $k$ limit is not a good approximation. In the CDM scenario, multi streaming effects, 
which are neglected in this framework, are certainly relevant for $k \agt 1\kMpc$ \cite{PueblasScoccimarro2009}. Moreover the separation of scales is
not so large for $k \lesssim 1\kMpc$. However, as for the propagator, a sensible choice is to  use an interpolation between small $k$ and large $k$ regimes. 
The equation derived in \cite{AnselmiPietroni2012} (see also Appendix \ref{eqPSappend}) takes the form
\bea
&&\partial_{\eta_a} \,\bar{P}_{ab}(k; \eta_a) = -\Omega_{ac}\, \bar{P}_{cb}(k; \eta_a)  -\Omega_{bc} \,\bar{ P}_{ac}(k; \eta_a)\nonumber \\
&&\;\; + H_{\bf{ a}}(k;\,  \eta_a,\eta_{in})\,  \bar{P}_{{\bf a}b}(k; \eta_{in})  +H_{\bf{ b}}(k;\, \eta_a,\eta_{in})\,  \bar{P}_{a{\bf b}}(k; \eta_{in})  \nonumber\\
&&\;\;+ \int \, ds'\;\big[\tilde{\Phi}_{ac}(k;\eta_a,s') \bar{G}_{bc}(k;\,\eta_a,s')+ \bar{G}_{ac}(k;\,\eta_a,s')\tilde{\Phi}_{cb}(k;s',\eta_a) \big],
\label{Tbar}
\eea    
where  $\tilde{\Phi}_{ac}$  is given by the weighted sum of the diagrams in Fig.~\ref{PHITilde}, and the explicit expression
generalized to our case of clustering DE is given by (see Appendix \ref{eqPSappend}),
\bea
\tilde{\Phi}_{ab}(k;\eta_a,\eta_b)&=&\exp\left[-\frac{1}{2}k^2 \sigma_v^2{\cal{I}}^2(\eta_a,\eta_b)\right]\left[{\Phi}_{ab}^{(1)}(k;\eta_a,\eta_b)\right.\nonumber\\
&+& \left. F(k) P_{in}(k) u_au_b (k^2\sigma_v^2)^2\frac{e^{\eta_a}}{C(\eta_a)}\frac{e^{\eta_b}}{C(\eta_b)}\, {\cal{I}}(\eta_a,\eta_{in})\,{\cal{I}}(\eta_b,\eta_{in})\right].\label{phitilde}
\eea  
Here  $\Phi_{ac}^{(1)}$ is the 1-loop approximation to $\Phi_{ac}$ given by (see the right panel of Fig. \ref{SigmaPhi})
\bea\label{Phi1loop}
&&\Phi_{ab}^{(1)}(k;\eta_a,\eta_b)= \nonumber\\
&&\qquad2 \frac{e^{\eta_a+\eta_b}}{C(\eta_a)C(\eta_b)} \int d^3 q \gamma_{acd}(\vec{k},-\vec{q},-\vec{p}) u_c P_{in}(q) u_e u_d P_{in}(p)u_f \gamma_{bef}(-\vec{k},\vec{q},\vec{p}).
\eea  
The second term of eq.~(\ref{phitilde}) contributes only for high $k$ values and it needs to be switched off for small values of the momenta, we do that by means of a filter function $F(k)$ defined as (see Appendix \ref{eqPSappend})
\be
F(k)=\frac{(k/\bar k)^4}{1+(k/\bar k)^4},\label{filter}
\ee  
where $\bar k$ is taken to be the scale $k$ at which the two contribution in eq.~(\ref{phitilde}) are equal at $z=0$ (which turns out to be of order $\bar k\simeq 0.2\kMpc$).
Note that with this $\tilde{\Phi}$ the 1-loop result is trivially recovered at small values of $k$ up to corrections of ${\cal O}(k^4)$.

In the rest of the paper we will consider eq.s~(\ref{dbargtot}) and (\ref{Tbar}), along with  (\ref{Ha}) and (\ref{phitilde}), and we 
solve them numerically \footnote{ In eq.  (\ref{Tbar}) one should  in principle use the propagator $\bar G$, obtained after solving eq. (\ref{dbargtot}). However, for the $\Lambda$CDM case it has been checked  that  the approximation of  replacing  $\bar G$ by $G^{eik}$ in that contribution   gives only sub percent differences  \cite{AnselmiPietroni2012} and the same happens in our case.
Therefore, for the sake of simplicity, we also consider this approximation here. }.

Notice that, as a by product of the resummation scheme and of the fact that $D_{m,c_s=0} \sim D_{c_s=1}$ at $1\%$ for viable models of quintessence, we expect the results for the power spectrum of the rescaled field $\varphi_1$ in clustering quintessence models to be very close to the corresponding one in smooth, $w$CDM models. This is one of the main results of this paper and it will be tested and exploited in the next section.

%%%%%%%%%%%%%%%%%%%%%%%%%%%%%%%%%%%%%%%%%%%%%%%%%%%%%%%%%%%%%%%%%%%%%%%%%%%%%%%%%%%%%%%%%%%%%%%%%
%%%%%%%%%%%%%%%%%%%%%%%%%%%%%%%%%%%%%%%%%%%%%%%%%%%%%%%%%%%%%%%%%%%%%%%%%%%%%%%%%%%%%%%%%%%%%%%%%
\section{Mapping smooth to clustering quintessence power spectra}
\label{results}

We now study the nonlinear solutions for the total density power spectrum both for smooth and clustering quintessence. We compare different models sharing the same values for the following cosmological parameters: $\Omega_{m,0}=0.25$, $\Omega_{b,0}h^2=0.0224$, $h=0.72$ and $n=0.97$. In addition, the amplitude of the initial power spectrum is also the same and corresponds to a $\sigma_{8}=0.8$ for the $\Lambda$CDM case $w=-1$. On the other hand, we consider different values of the DE equation of state parameter $w$, both for the extreme cases $c_s=0$ and $c_s=1$. In other terms, while the initial linear PS \cite{WellerLewis2003} is the same for all cosmologies, varying $w$ results in a different linear and nonlinear growth. In order to solve the PS evolution equations we set the initial conditions at redshift $z_{in}=1000$. This is done evolving back the $\Lambda$CDM linear PS from $z=0$ by means of linear theory in the Newtonian approximation. We verified that this prevents transient issues in the evolution 
equations.

In Appendix \ref{app:MICEcomp} we assess the range of validity of the approximation, comparing the results with measurements in N-body simulations for the $\Lambda$CDM model, since no simulations including DE perturbations are available at the moment. Our technique  reproduces at $1-2\%$ level the N-body power spectrum in the BAO regime, confirming previous results in \cite{AnselmiPietroni2012}. A $2\%$ accuracy extends to $k\sim 0.6\kMpc$ at $z=0.5$ and to  $k\sim 0.7\kMpc$ at higher redshifts. At $z=0$,  the  agreement with simulations is at the $2\%$ level only for $k \lesssim 0.25\kMpc$, and it worsens to the $5\%$ level up to $k\sim 0.6\kMpc$. This is, at some extent, expected given that we are working in the single-stream approximation. In \cite{PueblasScoccimarro2009}, the authors estimate the magnitude of corrections due to multi-streaming effects, which appears to be comparable to the discrepancy between the AP and N-body results\footnote{Notice that in \cite{PueblasScoccimarro2009} the authors 
assume $\sigma_{8}=0.9$: this high value enhances multi-streaming corrections.}.

% FIG 5
\begin{figure}[t!]
\begin{center}
{\includegraphics[width=0.6\textwidth]{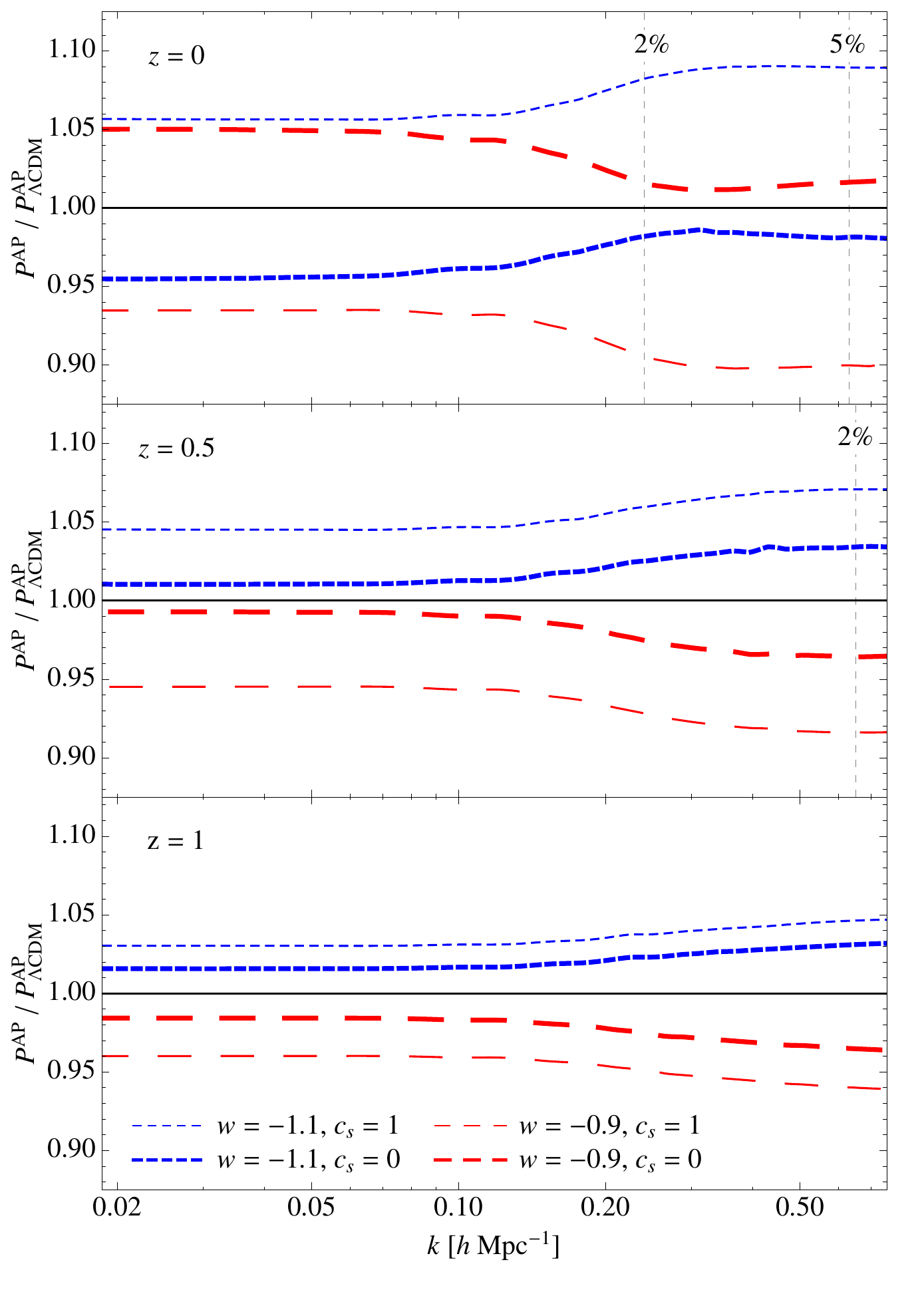}}
\caption{\small{Ratio of the AP predictions for the total nonlinear power spectrum in different DE models to the $\Lambda$CDM one at  $z=0$, $z=0.5$ and $z=1$ ({\em top to bottom}). We consider, in particular, the cases of $w=-0.9$ ({\em long-dashed, red curves}) and $w=-1.1$ ({\em short-dashed, blue}) both for $c_s=1$ ({\em thin curves}) and $c_s=0$ ({\em thick curves}). The vertical dashed lines mark the limit of validity of the AP resummation (see Appendix~\ref{app:MICEcomp}).}}
\label{PSoverLCDM}
\end{center}
\end{figure}

We turn now to DE effects. We are mainly interested in the {\em relative difference} of the AP predictions for the distinct models considered. One could expect (needless to say, without guarantee) that predictions for {\em relative} DE effects, possibly comparable to or smaller than  the method accuracy, are more robust since overall systematic errors could be reduced.   
Fig.~\ref{PSoverLCDM} shows the ratio of AP prediction for different DE scenarios to the one for the corresponding $\Lambda$CDM cosmology. We consider, in particular, the cases of $w=-0.9$ ({\em long-dashed, red curves}) and $w=-1.1$ ({\em short-dashed, blue}) both for $c_s=1$ ({\em thin curves}) and $c_s=0$ ({\em thick curves}). In all cases we show the {\em total} perturbations power spectrum, coinciding with the matter one for $c_s=1$. Panels from top to bottom show the outcomes at redshifts $z=0$, $z=0.5$ and $z=1$. The vertical dashed lines indicate the limits of the resummation as explained above.

A close look at the plots reveals that the power spectrum differences between smooth and clustering DE cosmologies are approximately given by multiplicative factors.  These results confirm the behavior already underlined by \cite{DAmicoSefusatti2011} and \cite{SefusattiVernizzi2011}:  both at linear and nonlinear level for $w>-1$ ($w<-1$) the total perturbations for $c_s=0$ are enhanced (suppressed) w.r.t. the matter ones for $c_s=1$. In particular, for $c_s=0$, the relative nonlinear corrections for $w>-1$ ($w<-1$) are smaller (larger) than in the $\Lambda$CDM case.

% FIG 6
\begin{figure}[t!]
\begin{center}
{\includegraphics[width=0.6\textwidth]{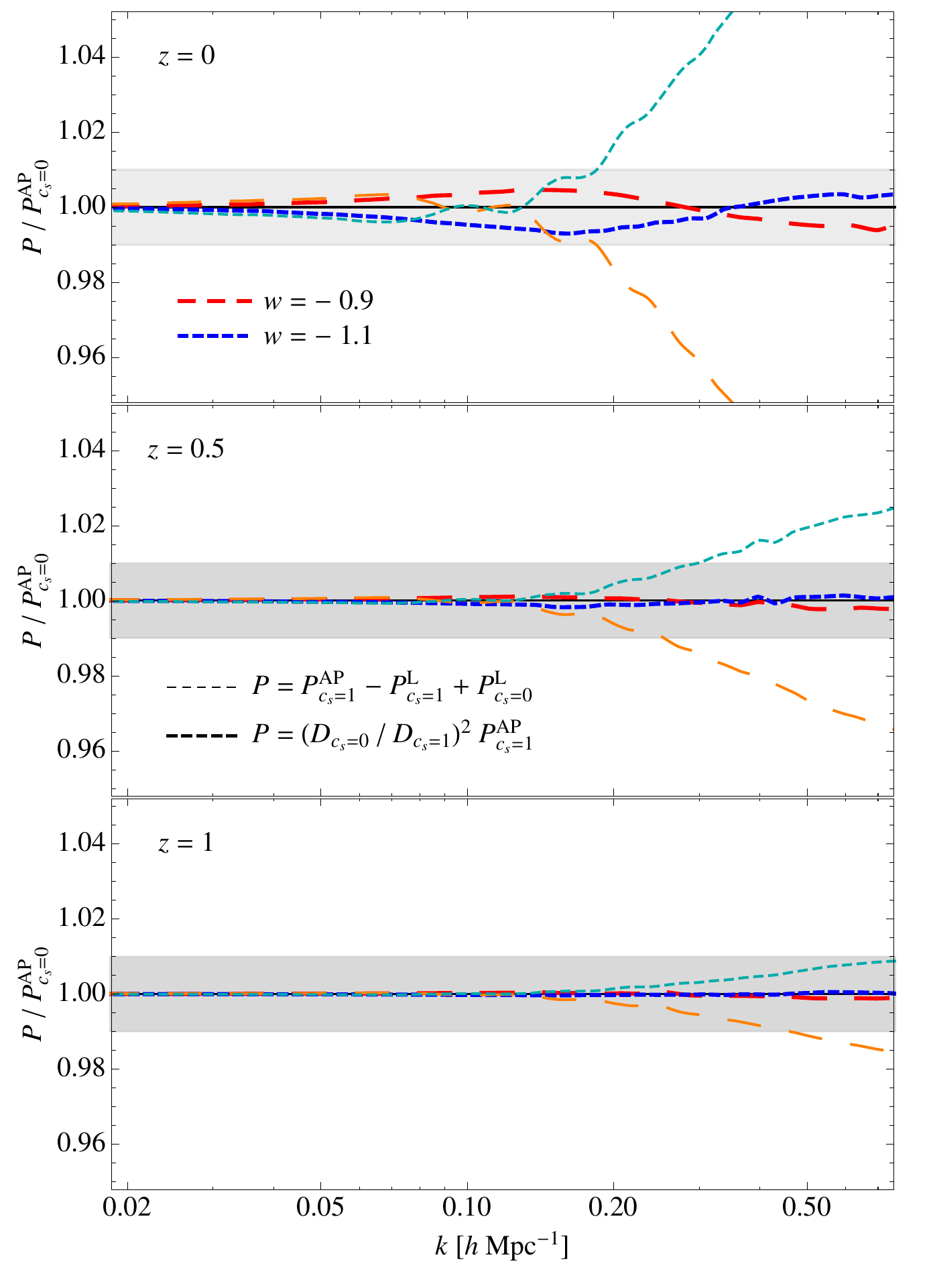}}
\caption{\small{Ratio of the total, $c_s=0$, nonlinear power spectrum computed with approximations of eq.s~(\ref{linApp}) ({\em thin curves}) and (\ref{nlApp}) ({\em thick curves}) to the full AP prediction, $P_{c_s=0}^{AP}$, for 
$w=-0.9$ ({\em long-dashed curves}) and $w=-1.1$ ({\em short-dashed}). }}
\label{figTestQC}
\end{center}
\end{figure}

It is worth investigating in greater detail the relation between the nonlinear clustering in smooth and clustering DE scenarios. Ref. \cite{AnselmiBallesterosPietroni2011} found that the absolute value of nonlinear corrections, for the total power spectrum, in the $c_s=0$ and $c_s=1$ models is comparable, at better than $1\%$ in the BAO region. This corresponds to the approximation
\be
P_{c_s=0}(k; \eta)-P_{c_s=0}^{L}(k; \eta) \simeq P_{c_s=1}(k; \eta)-P_{c_s=1}^{L}(k; \eta)\, .
\label{linApp}
 \ee
This implies that, when this approximation holds, the effects of quintessence on the nonlinear evolution are essentially those induced by the background evolution, which depends only on the equation of state parameter $w$, and are therefore practically independent on the speed of sound $c_s$. On the other hand the analytical results described in Sections \ref{dynamics} and \ref{StatisticalObs} suggest a different relation, that is 
\be\label{nlApp}
P_{c_s=0}(k; \eta) \simeq \left[ \frac{D(\eta)_{c_s=0}}{D(\eta)_{c_s=1}}\right]^{2} P_{c_s=1}(k; \eta)\,.
\ee
In Fig.~\ref{figTestQC} we compare both approximations of eq.s~(\ref{linApp}) and (\ref{nlApp}) considering their ratio to the exact AP prediction, $P_{c_s=0}^{AP}(k; \eta)$, assumed as reference, for redshifts $z=0$, $0.5$ and $1$. For (\ref{nlApp}) we use thick lines  and the same legends as  before ( i.e.,  {\em long-dashed, red curves} for $w=-0.9$ and {\em short-dashed, blue curves } for $w=-1.1$),  while  for the linear assumption (\ref{linApp}) we use thin lines
({\em long-dashed, orange curves} for $w=-0.9$ and {\em short-dashed, green curves } for $w=-1.1$). 

We can see that the approximation of eq.~(\ref{nlApp}) works at $1\%$ level at $z=0$ and even better for $z>0$ for corrections due to $\Delta w\sim 0.1$ variations. The linear assumption of eq.~(\ref{linApp}) confirms its validity on BAO scales, however, for larger values of $k$ it abruptly breaks down. This suggests that, even tough quintessence perturbations are smaller w.r.t. matter ones, their interaction with the latter is quite significant at small scales.  

An immediate consequence of the validity of eq.~(\ref{nlApp}) is that it provides a way to compute an approximation to the total power spectrum $P_{c_s=0}(k)$ taking advantage of numerical predictions for the matter power spectrum for more standard ($c_s=1$) quintessence model, including emulators such as the ones of \cite{HeitmannEtal2014, AgarwalEtal2013} or the  \texttt{halofit} fitting function \cite{SmithEtal2003, TakahashiEtal2012}. This is quite an interesting result since no proper N-body results is yet available for cosmologies characterized by dark energy perturbations.

\begin{figure}[t!]
\begin{center}
{\includegraphics[width=0.6\textwidth]{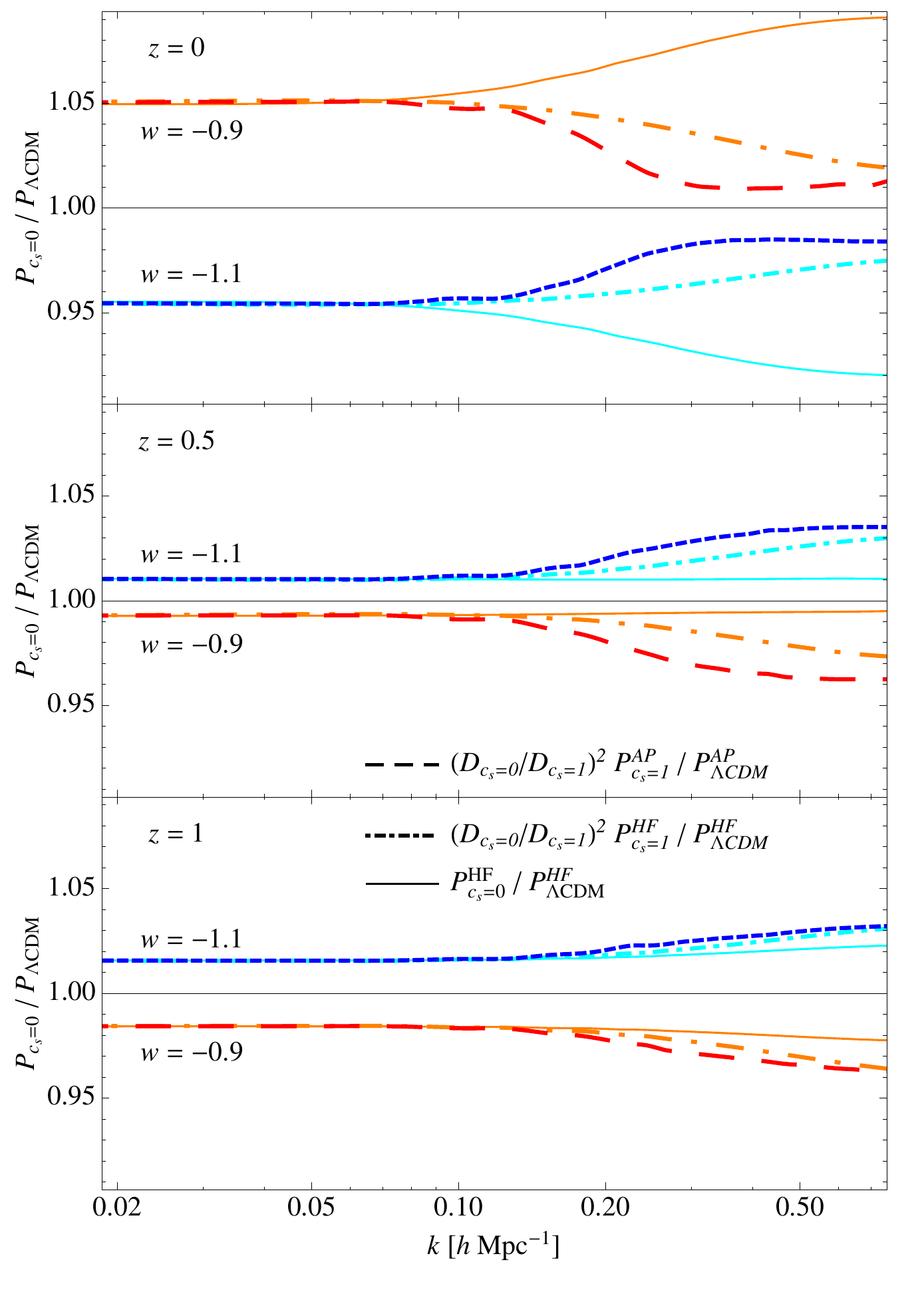}}
\caption{\small{Ratio of the total power spectrum $P_{c_s=0}(k)$ to the $\Lambda$CDM prediction computed in the approximation of eq.~(\ref{nlApp}) using the AP resummation ({\em dashed curves}) and the  \texttt{halofit} (HF) formula ({\em dot-dashed curves}) for $w=-0.9$ and $w=-1.1$, at $z=0$, $0.5$ and $1$ ({\em panels top to bottom}). Also shown is the prediction assumed in \cite{AmendolaKunzSapone2008} ({\em continuous curves}).}}
\label{AP_HF}
\end{center}
\end{figure}

Fig.~\ref{AP_HF} shows the ratio of the total power spectrum $P_{c_s=0}(k)$ to the $\Lambda$CDM prediction computed in the approximation of eq.~(\ref{nlApp}) using the AP resummation ({\em dashed curves}) and the  \texttt{halofit} formula ({\em dot-dashed curves}) at $z=0$, $0.5$ and $1$ ({\em panels top to bottom}). We consider models characterized by $w=-0.9$ ({\em long dashed and dot-dashed}) and $w=-1.1$ ({\em short dashed and dot-dashed}). We notice a rather large discrepancy between the AP and  \texttt{halofit} prediction for the relative effect around scales of order $k\sim 0.3\kMpc$ that might be related, in part, to the 5\% accuracy of the  \texttt{halofit} formula. Still, at smaller scales such discrepancy is reduced and the two approach seem to reach a possible, common asymptotic value. Despite these caveats,  \texttt{halofit} could provide, within certain limits, a good tool to approximately describe DE clustering, for instance in the context of surveys forecasts. In this respect, we notice that 
until now, since no clear theoretical indication on how to describe the nonlinear evolution of DE perturbations was available, forecasts for the detectability of this kind of departures from a $\Lambda$CDM cosmology, relied on rather arbitrary assumptions. 
For instance, ref.~\cite{AmendolaKunzSapone2008} considered the  \texttt{halofit} formula as a generic and direct mapping ${\mathcal F}$ between the linear and nonlinear power spectra, i.e. $P(k)={\mathcal F}_{HF}[P_L(k)]$. In the case of clustering quintessence this amounts to the prediction
\be\label{eq:amendola}
P_{c_s=0}(k)={\mathcal F}_{HF}[P_{c_s=0}^L(k)]\,,
\ee
with is clearly quite different from eq.~(\ref{nlApp}). Such prediction, denoted as $P_{c_s=0}^{HF}(k)$ is shown as a continuous curve in Fig.~\ref{AP_HF}, where we notice that it leads to the opposite relative effect at low redshift, and almost no additional nonlinear correction at higher $z$. Since no particular theoretical motivation could be given for eq.~(\ref{eq:amendola}), ref.~\cite{AmendolaKunzSapone2008} also provides a comparison with the expected constraints under the assumption of eq.~(\ref{nlApp}) finding larger errors. However, it is important to point out that the two assumptions lead to quite different effects, under a phenomenological point of view.  

A final confirmation of the validity of eq.~(\ref{nlApp}) and a quantification of the systematic errors it entails will require a test based on numerical simulations. For the time being, however, we can venture to suggest a plausible analytical interpolation for the total power spectrum of models of quintessence when $0 < c_s <1$. In fact, we do not known how to solve the nonlinear fluid equations for a generic speed of sound, since, in the first place, the matter and DE fluids are not comoving. As a further complication, already at linear level, a new fundamental scale, the sound horizon of quintessence perturbations, enters the game leading to a scale-dependent linear growth function. However, $c_s=0$ represents a limiting case for these models whose properties, in general, are expected to interpolate between the extreme values $c_s=0$ and 1.  Indeed, at linear level, for scales larger than the sound horizon, the quintessence fluid clusters while, at smaller scales, perturbations are erased. In both these 
regimes eq.~(\ref{nlApp}) applies, allowing us to promote it to any value of $k$ for models with generic speed of sound $c_s$ as
\be
P_{c_s}(k; \eta) \simeq \left[ \frac{D(k; \eta)_{c_s}}{D(\eta)_{c_s=1}}\right]^{2} P_{c_s=1}(k; \eta)\, .
\label{nlApp_cs}
\ee
We stress again that our analysis relies on the single-stream approximation, and nothing is known about the nonlinear behavior of DE clustering when multi-streaming effects become dominant. Still, we believe that our results are an important step toward theoretically-motivated predictions for the nonlinear density power spectrum in clustering quintessence models.

%%%%%%%%%%%%%%%%%%%%%%%%%%%%%%%%%%%%%%%%%%%%%%%%%%%%%%%%%%%%%%%%%%%%%%%%%%%%%%%%%%%%%%%%%%%%%%%%%
%%%%%%%%%%%%%%%%%%%%%%%%%%%%%%%%%%%%%%%%%%%%%%%%%%%%%%%%%%%%%%%%%%%%%%%%%%%%%%%%%%%%%%%%%%%%%%%%%
\section{Conclusions}
\label{discussion}

Regardless any proper theoretical motivation or degree of naturalness, quintessence models characterized by a vanishing speed of sound are extremely interesting under a phenomenological point of view. In this case, in fact, quintessence perturbations are present at all scales and, more importantly, in the single-stream approximation, they are comoving with perturbations in the matter component, i.e.~they share the same velocity field. This allows to easily extend sophisticated techniques in cosmological perturbation theory already developed for standard, $\Lambda$CDM cosmologies, \cite{SefusattiVernizzi2011}.

At the same time, $c_s=0$ represents a limiting case of a much broader class of quintessence models,  parametrized by the equation of state parameter $w$ (with allowed values close to $-1$) and the  sound speed parameter $c_s$ taking values, in principle, in the relatively wide interval $0\le c_s\le1$. Needless to say, the possible detection of any departure from the standard values $w=-1$ and $c_s=1$ would be of great significance toward an understanding of the mechanism driving the observed acceleration of the Universe. Still, the theoretical description of the nonlinear evolution of quintessence perturbation poses a remarkable challenge. The difficulty of the task is highlighted by the fact that no numerical simulation for these quintessence models has been performed to date. Furthermore, it is easy to imagine that when they {\em will} be performed they will likely involve relevant approximations. Precisely for this reason any analytical insight into the nonlinear behavior of a central quantity as the 
density power spectrum is greatly valuable. Approximate solutions for the power spectrum of (standard) models with $c_s=1$ and (extreme) models with a vanishing speed of sound will therefore mark the limits of the broader set of solutions for models with intermediate values of $c_s$ and provide boundaries for useful interpolations, both under a practical as under a theoretical point of view.

With this goal in mind, we applied, in this paper, the resummation scheme proposed in \cite{AnselmiPietroni2012} to quintessence models with $c_s=0$, providing predictions for the total (i.e. matter and quintessence) density power spectrum, in the standard single-fluid approximation. This approach extends the range of validity of perturbative solutions beyond the acoustic oscillations scales up to $k\sim 0.6\kMpc$ with an accuracy below the 2\% level at redshifts $z\ge 0.5$, therefore comparable to the accuracy delivered by the  \texttt{halofit} fitting formula \cite{SmithEtal2003, TakahashiEtal2012} or the emulator approach of \cite{HeitmannEtal2014}.

In addition, our analytical investigation suggests an approximate but useful and simple mapping between the nonlinear power spectra of models with $c_s=1$ and models with $c_s=0$ when all other cosmological parameters are held at the same values. We show, in fact, that the ratio between these two quantities is very close to the ratio of the square of their respective linear growth factors, eq.~(\ref{nlApp}). This is the likely consequence of the fact, already suggested in earlier works \cite{SefusattiVernizzi2011, AnselmiBallesterosPietroni2011, DAmicoSefusatti2011}, that nonlinear growth is driven predominantly by the matter component. 

This is quite an interesting result as it provides, for the first time, a theoretical motivation for simple predictions of the density power spectrum that can make use of tools as  \texttt{halofit} or more in general results from existing, numerical simulations of more standard models. Waiting for a proper N-body description of clustering quintessence models, our findings will enable more robust assessments of the ability of future cosmological probes to constrain this class of models. We will address these issues in future work.

\acknowledgments

We thank  M. Crocce  and  M. Pietroni for their careful reading of the manuscript and comments. We are grateful to A. Taruya and T. Nishimichi for timely providing us with a helpful implementation of their revised \texttt{halofit} formula, and to  the MICE collaboration for making available to us the power spectrum measurements in their recent Grand Challenge simulation. S.A. is thankful the Abdus Salam ICTP for support and hospitality when this work was initiated. E.S. and S.A. have been supported in part by NSF-AST 0908241.

\appendix

%%%%%%%%%%%%%%%%%%%%%%%%%%%%%%%%%%%%%%%%%%%%%%%%%%%%%%%%%%%%%%%%%%%%%%%%%%%%%%%%%%%%%%%%%%%%%%%%%
%%%%%%%%%%%%%%%%%%%%%%%%%%%%%%%%%%%%%%%%%%%%%%%%%%%%%%%%%%%%%%%%%%%%%%%%%%%%%%%%%%%%%%%%%%%%%%%%%
\section{Nonlinear propagator and power spectrum in the large $k$ or eikonal limit}\label{appLargek}

In this appendix we present some details of the calculation of the propagator, $G_{ab}^{eik}(k,\eta_a,\eta_b)$, and the power spectrum, 
$P_{ab}^{eik}(k,\eta_a,\eta_b)$, in the eikonal limit.
\begin{figure}[t]
\begin{center}
\includegraphics[width=0.98\textwidth]{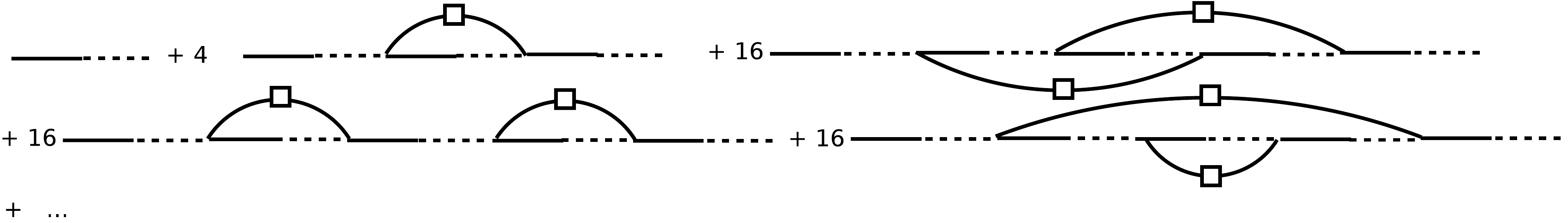}
\end{center}
\caption{ Chain-diagrams contributing to the propagator  $-i\,G_{ab}^{eik}(k,\eta_a,\eta_b)$ up to two loops, where the dots stand for similar chain-diagrams with more loops.}
\label{FigA2}
\end{figure}

For the  propagator we need to sum the chain-diagrams shown in Fig.~\ref{FigA2}, using the Feynman rules of Fig.~\ref{FeynmanRules}. Since the rules are exactly the same as those for the standard case with the only exception that for each vertex we need to include a time-dependent factor of $C^{-1}$ evaluated at the appropriate time, we restrict here to point out where exactly such factors enter. Following the computation of \cite{CrocceScoccimarro2006B}, the sum of such diagrams can be written as 
\be\nonumber
 \,G_{ab}^{eik}(k,\eta_a,\eta_b)=\,g_{ab}(\eta_a,\eta_b)\sum_{n=0}^{+\infty}(2n-1)!! (-k^2 \sigma_v^2)^n\int_{\eta_b}^{\eta_a}ds_1\int_{\eta_b}^{s_1}ds_2 ...\int_{\eta_b}^{s_{2n-1}} ds_{2n} \prod_{j=1}^{2n}\frac{e^{s_{j}}}{C(s_{j})},
 \ee 
 where $\sigma_v$ is given by eq.~(\ref{sigmav}). Here, the factor $(2n-1)!!$ accounts for all the possible chain-diagrams with $n$ loops, and we have used the composition  property of the linear propagator $\,g_{ab}(\eta_a,\eta_b)=g_{ac}(\eta_a,s)g_{cb}(s,\eta_b)$. Note that the symmetry factor $2^{2n}$ corresponding to an $n$-loop diagram cancel with  the factor  $1/2^{2n}$  coming from the presence of $2n$ vertexes, which in the large $k$ limit contribute as 
\be u_c\gamma_{acb}(\vec{k},\vec{q}-\vec{k},-\vec{q})\to\frac{1}{2}\frac{\vec{k}\cdot\vec{q}}{q}\delta_{ab}.
\ee  
To obtain the final result we need to perform the time integrals, which in our case cannot be done analytically. However,  as the integrand is symmetric in all the variables we can use the property
\bea 
\int_{\eta_b}^{\eta_a}ds_1\int_{\eta_b}^{s_1}ds_2 ...\int_{\eta_b}^{s_{2n-1}} ds_{2n} \prod_{j=1}^{2n}\frac{e^{s_{j}}}{C(s_{j})} & = & \frac{1}{(2n)!}\int_{\eta_b}^{\eta_a}ds_1\int_{\eta_b}^{\eta_a}ds_2 ...\int_{\eta_b}^{\eta_a} ds_{2n} \prod_{j=1}^{2n}\frac{e^{s_{j}}}{C(s_{j})}\nonumber\\
&= &\frac{1}{(2n)!}\left[\int_{\eta_b}^{\eta_a}ds \frac{e^{s}}{C(s)}\right]^{2n}\equiv\frac{{\cal{I}}^{2n}(\eta_a,\eta_b)}{(2n)!}
\label{proper}
\eea
 to rewrite the propagator as
 \bea
 \,G_{ab}^{eik}(k,\eta_a,\eta_b)&=&\,g_{ab}(\eta_a,\eta_b)\sum_{n=0}^{+\infty}\frac{(2n-1)!!}{(2n)!} (-k^2 \sigma_v^2{\cal{I}}^2(\eta_a,\eta_b))^n\nonumber\\
 &=&\,g_{ab}(\eta_a,\eta_b)\exp\left[-\frac{1}{2}k^2 \sigma_v^2{\cal{I}}^2(\eta_a,\eta_b)\right]. \label{propeik}
 \eea  which corresponds to the  r.h.s. of eq.~(\ref{Glargek1}).

\begin{figure}[t]
\begin{center}
\includegraphics[width=0.9\textwidth]{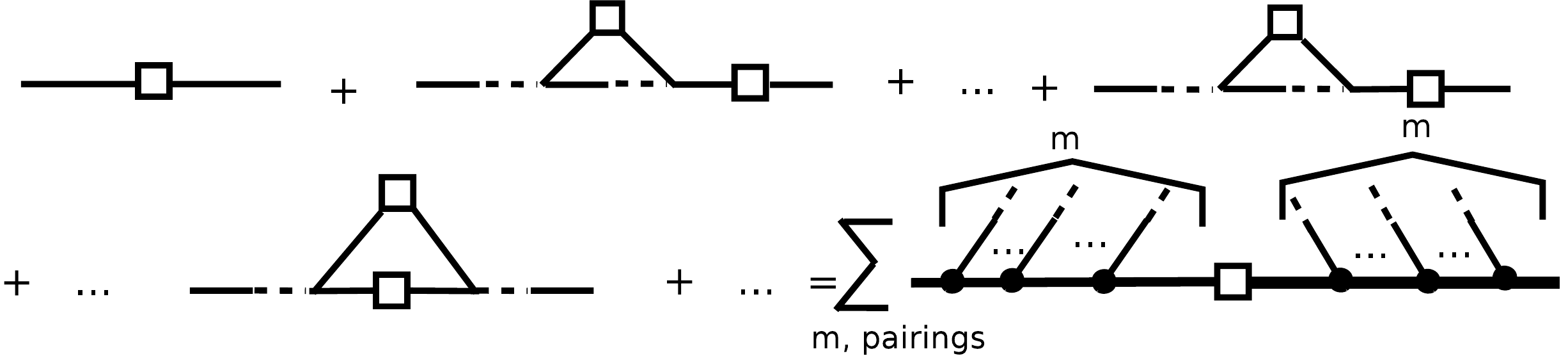}
\end{center}
\caption{ Diagrams contributing to  $P_{ab}^{eik}$, where the thick lines  in the last diagram are associated to  $-i G_{ab}^{eik}$ given in eq. (\ref{propeik})  and the big dots correspond  to renormalized vertexes $\Gamma_{dce}^{eik}$ defined in eq. (\ref{renvert}).}
\label{pseik}\end{figure}

The diagrams that dominate the large $k$ limit of the power spectrum are those in which all corrections involving the PS evaluated in soft momenta $q\ll 1$ are connected with a ``hard'' line carrying momenta of order $k$. Some of the diagrams are shown in Fig. \ref{pseik}, where we see that some of them involve a correction to the left side of the hard power spectrum, to the right side, while others are joining the two sides, and we need to sum all of them together.  It is not difficult to see that the sum of all the diagrams involving  correction only to the left side of the hard power spectrum plus the ones that only involve correction to the right side yield, simply, 
\be 
G_{ac}^{eik}(k,\eta_a,\eta_{in})\,G_{bd}^{eik}(k,\eta_b,\eta_{in})u_cu_d P_{in}(k).
\ee 
Now, if we consider the same sum but now with a soft PS that  joins the two sides of the hard PS, then, each term in the previous sum  contains two  additional  vertices (one on the left and one on the right) evaluated at given times. For instance, for the one on the right we have 
\bea
&& g_{ad}(\eta,s) e^s \gamma_{dce}(\vec{k},-\vec{q},\vec{q}-\vec{k}) u_c\, g_{eb}(s,\eta')\nonumber\\
&&\qquad \to g_{ab}(\eta,\eta')\frac{ e^s}{C(s)}\frac{1}{2} \frac{\vec{k} \cdot \vec{q}}{q^2 }\,\Theta(\eta-s)\Theta(s-\eta') \qquad\qquad ({\rm for}\; k\gg q)\,,
\label{vtreegg}
\eea where we use that $\,g_{ab}(\eta,\eta')=g_{ad}(\eta,s)g_{db}(s,\eta')$ and we have included the $u_c$ that comes from the  soft PS which is to be joined to the soft  vertex line.
 From this equation we see that, in the large $k$ limit, the time integrations corresponding to the vertex insertions factorize from the ones correcting the propagator $g_{ab}(\eta,\eta')$ on the hard line.  Then,  analogously to the above computation of $G^{eik}$,  we can conclude that  the renormalized version of eq. (\ref{vtreegg}), obtained  by including all possible soft PS insertions on the hard line, can be written as  
\bea
 &&\int ds_1 \int ds_2\,G^{eik}_{ad}(k,\eta,s_1)\Gamma_{dce}^{eik}(\vec{k},-\vec{q},\vec{q}-\vec{k},s_1,s,s_2)  G^{eik}_{ad}(|\vec{q}-\vec{k}|,\eta,s_1)\nonumber\\
 &&\to \,G^{eik}_{ad}(k,\eta,\eta') \frac{e^{s}}{C(s)}\frac{1}{2}\frac{\vec{k}\cdot\vec{q}}{q^2}\Theta(\eta-s)\Theta(s-\eta') \qquad\qquad ({\rm for}\; k\gg q)\,, 
\label{renvert}
\eea 
where the notation $\Gamma_{dce}^{eik}$ denotes a renormalized vertex. 

Therefore,  to compute the sum of all terms with  the above correction, we just need to  contract the two additional vertexes (i.e. to multiply  by a soft PS and  to integrate over the soft momenta), obtaining
\bea
&&\,G_{ac}^{eik}(k,\eta_a,\eta_{in})\,G_{bd}^{eik}(k,\eta_b,\eta_{in})u_cu_d P_{in}(k)\int_{\eta_{in}}^{\eta_a}ds_r\int_{\eta_{in}}^{\eta_b}ds_l\frac{ e^{s_r+s_l}}{C(s_r)C(s_l)}\nonumber\\
&&\times\,\Theta(\eta_a-s_r)\Theta(s_r-\eta_{in})\Theta(\eta_{b}-s_l)\Theta(s_l-\eta_{in}) \int d^3q \frac{(\vec{k} \cdot \vec{q})^2}{q^4 } P_{in}(q)\nonumber\\
&&=\,G_{ac}^{eik}(k,\eta_a,\eta_{in})\,G_{bd}^{eik}(k,\eta_b,\eta_{in})u_cu_d P_{in}(k)\int_{\eta_{in}}^{\eta_a}ds_r\int_{\eta_{in}}^{\eta_b}ds_l\frac{ e^{s_r+s_l}}{C(s_r)C(s_l)} (k^2\sigma_v^2),
\eea where we added the symmetry factor.  Notice that, due to the presence of the step functions, the time integrations corresponding to the vertex on the right side of the  hard PS  should  be causally ordered separately from those of the left. 

One way of organizing the full sum is to  group all terms with $m$  soft PS  joining the two sides of the hard PS, and then perform the sum over $m$. This procedure is illustrated in Fig.~\ref{pseik}, where the thick lines correspond to renormalized propagators $\,G^{eik}$,  and the thick dots represent renormalized vertices  $\Gamma_{dce}^{eik}$. Doing this, we get
\bea
P_{ab}^{eik}(k,\eta_a,\eta_b)&=&\,G_{ac}^{eik}(k,\eta_a,\eta_{in})\,G_{bd}^{eik}(k,\eta_b,\eta_{in})u_cu_d P_{in}(k)\nonumber\\
&\times&\sum_{m=0}^{+\infty}\int_{\eta_{in}}^{\eta_a}ds_{r1}...\int_{\eta_{in}}^{s_{r(m-1)}}ds_{rm} \prod_{rj=0}^{m}\frac{e^{s_{rj}}}{C(s_{rj})}\nonumber\\
&\times&\int_{\eta_{in}}^{\eta_b}ds_{l(m-1)}...\int_{\eta_{in}}^{\eta_b}ds_{lm} \prod_{lj=0}^{m}\frac{e^{s_{lj}}}{C(s_{lj})}, (k^2\sigma_v^2)^m m!,
\eea where the factor $m!$ is the number of different ways of pairing the additional vertexes.   
 Note that it is only  in the large $k$ limit that the time integrations corresponding to the vertex insertions  decouple from those correcting the linear propagator yielding only one renormalized propagator $\,G_{bd}^{eik}(k,\eta_{a,b},\eta_{in})$ for each side of the hard PS. 
 
Finally, using the property of eq.~(\ref{proper}), we obtain
\bea
P_{ab}^{eik}(k,\eta_a,\eta_b)&=&\,G_{ac}^{eik}(k,\eta_a,\eta_{in})\,G_{bd}^{eik}(k,\eta_b,\eta_{in})u_cu_d P_{in}(k) \sum_{m=0}^{+\infty}\frac{1}{m!}\left[ k^2\sigma_v^2{\cal{I}}(\eta_a,\eta_{in}){\cal{I}}(\eta_b,\eta_{in})\right]^m\nonumber\\
&=&\,G_{ac}^{eik}(k,\eta_a,\eta_{in})\,G_{bd}^{eik}(k,\eta_b,\eta_{in})u_cu_d P_{in}(k) \exp\left[ k^2\sigma_v^2{\cal{I}}(\eta_a,\eta_{in}){\cal{I}}(\eta_b,\eta_{in})\right]\nonumber\\
&=& P_{in}(k) u_a u_b \exp\left[-\frac{1}{2} k^2\sigma_v^2{\cal{I}}^2(\eta_a,\eta_b)\right],\label{PSeik}
\eea where in the last equality we have used eq.~(\ref{propeik}).

%%%%%%%%%%%%%%%%%%%%%%%%%%%%%%%%%%%%%%%%%%%%%%%%%%%%%%%%%%%%%%%%%%%%%%%%%%%%%%%%%%%%%%%%%%%%%%%%%
\section{Evolution equations: factorization and interpolation at intermediate scales}\label{EvolutEq}

\subsection{Factorization in the equation for the propagator} \label{appFact}

Now that we have the propagator in the large $k$ limit, it is straightforward  to show that the factorization in eq.~(\ref{TRGL}) applies in this limit. To check the factorization,  we can start by using the solution for the propagator of eq.~(\ref{propeik}) to write the r.h.s. of eq.~(\ref{dgtot}) as
\bea
\partial_{\eta_a}\, G_{ab} ^{eik}(k;\, \eta_a, \eta_b)
& = &\partial_{\eta_a}\, g_{ab} (k;\, \eta_a, \eta_b) \exp\left[-\frac{1}{2}k^2 \sigma_v^2{\cal{I}}^2(\eta_a,\eta_b)\right]
\nonumber\\&  &
+ g_{ab} (k;\, \eta_a, \eta_b) \exp\left[-\frac{1}{2}k^2 \sigma_v^2{\cal{I}}^2(\eta_b,\eta_a)\right]\left[-k^2 \sigma_v^2{\cal{I}}(\eta_a,\eta_b)\right] \frac{e^{\eta_a}}{C(\eta_a)}\nonumber\\ 
& = & \delta_{ab}\delta_D(\eta_a-\eta_b)-\Omega_{ac}G_{cb}^{eik}(\eta_a,\eta_b)
\nonumber\\&  &
+G_{ab} ^{eik}(k;\, \eta_a, \eta_b)\left[-k^2 \sigma_v^2{\cal{I}}(\eta_a,\eta_b)\right] \frac{e^{\eta_a}}{C(\eta_a)}
\eea 
where in the last equality we have used that the linear propagator satisfies eq.~(\ref{eqlinprop}).
Then, by comparing this to eq.~(\ref{dgtot}) we can read
\bea
\Delta G_{ac}^{eik}(k;\,\eta_a,\eta_b) & = & H_{{\bf a}}^{eik}(k;\,\eta_a,\eta_b)\, G_{{\bf a}c}^{eik}(k;\,\eta_a,\eta_b)\nonumber\\
& = & G_{ab} ^{eik}(k;\, \eta_a, \eta_b)\left[-k^2 \sigma_v^2{\cal{I}}(\eta_a,\eta_b)\right] \frac{e^{\eta_a}}{C(\eta_a)}\, ,
\label{TRGL2}
\eea 
where the bold indices are not summed up, hence 
\be\label{heik}
H_{{a}}^{eik}(k;\, \eta_a,\eta_b) =u_{ a}\left[-k^2 \sigma_v^2{\cal{I}}(\eta_a,\eta_b)\right]\frac{e^{\eta_a}}{C(\eta_a)}.
\ee
On the other hand, the 1-loop approximation to $\Sigma_{ad}$ is given in eq. (\ref{sigform}), and in the large $k$ limit  we have
   \be
\int_{\eta_b}^{\eta_a}ds\,\Sigma_{ab}^{eik(1)}(k;\eta_a,s)u_{ b} =u_{a}\left[-k^2 \sigma_v^2{\cal{I}}(\eta_a,\eta_b)\right] \frac{e^{\eta_a}}{C(\eta_a)},
\ee  
and then, from the last two equation, we conclude that
   \be
\int_{\eta_b}^{\eta_a}ds\,\Sigma_{ab}^{eik(1)}(k;\eta_a,s)u_{ b} =H_{{ a}}^{eik}(k;\, \eta,s),
\ee  
which is the result we wanted to show.

\begin{figure}[t]
\begin{center}
\includegraphics[width=0.9\textwidth]{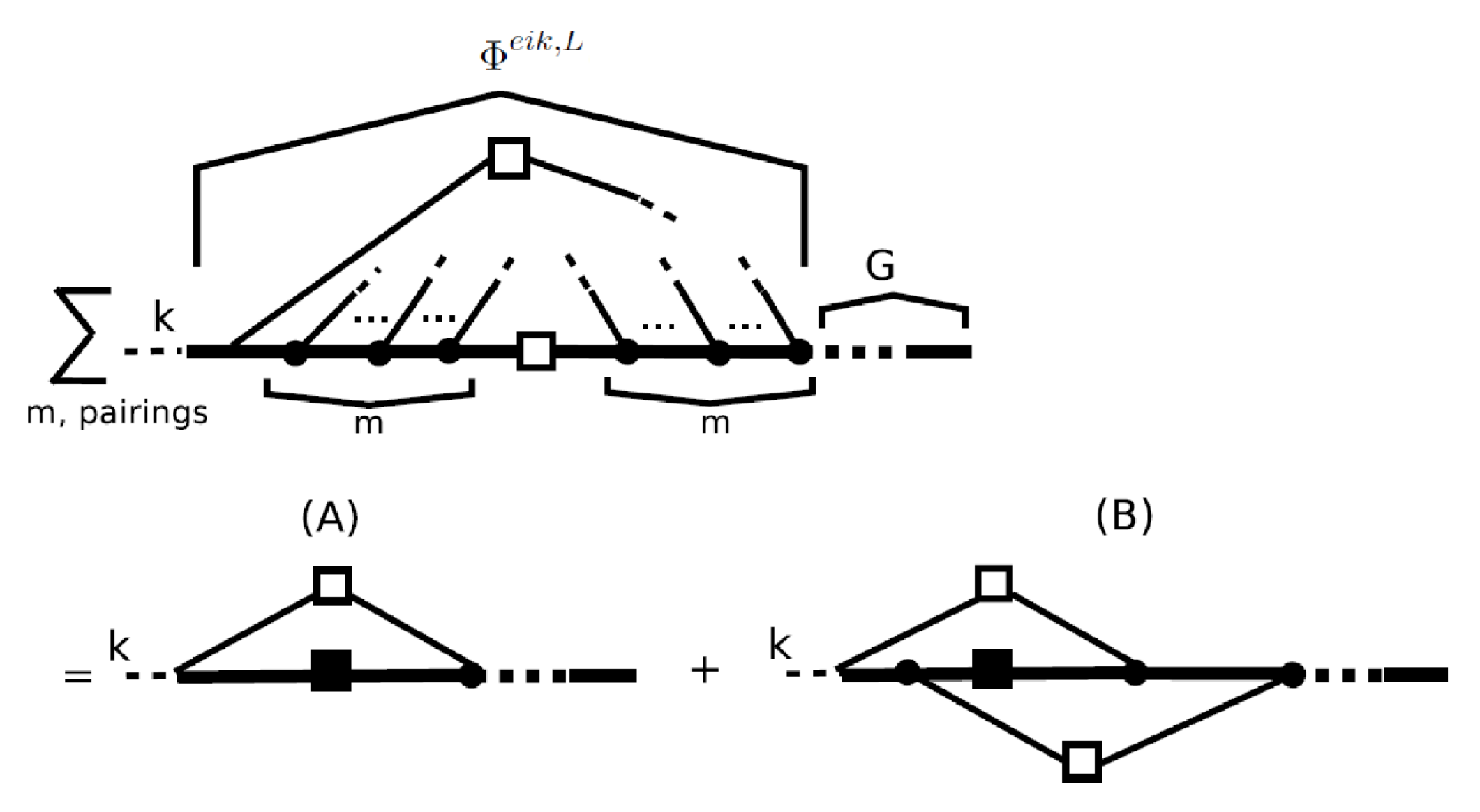}
\end{center}
\caption{Diagrams contributing to $\Phi_{ad}^{eik,L}(k;  \eta,s) G_{bd}^{eik}(k;\eta,s)$, where  thick lines,  thick dots, and   black PS represent respectively, $G^{eik}$  (given in eq. (\ref{propeik})), renormalized vertexes $\Gamma_{dce}^{eik}$ (defined in eq. (\ref{renvert}), and $P^{eik}$   (given by eq. (\ref{PSeik})).}
\label{PHIG}
\end{figure}

\subsection{Resummation and  evolution equation  of the  power spectrum}\label{eqPSappend}

Let us  now consider the evolution equation of the power spectrum. Following Ref.~\cite{AnselmiPietroni2012},  we provide here some details
of the computations that motivate the use of the approximation given in eq.~(\ref{Tbar}) to the exact evolution equation eq.~(\ref{Texact}).

As for the propagator, the approximated evolution equation is obtained as an interpolation between those corresponding to  the large $k$ and small $k$ regimes.
On the one hand,  for small values of $k$, eq.~(\ref{Texact}) can be approximated at 1-loop level, yielding
\bea
&&\partial_{\eta_a} \,P_{ab}^{(1)}(k; \eta_a) = -\Omega_{ac} \,P_{cb}^{(1)}(k; \eta_a)  -\Omega_{bc} \,P^{(1)}_{ac}(k; \eta_a)
\nonumber \\ &&\;\;  
+P_{in}(k)\lp H_{{ a}}(k;\, \eta_a,\eta_{in})u_b+H_{{ b}}(k;\, \eta_a,\eta_{in}u_a)  \rp
\nonumber\\
&&\;\;+ \int \, ds'\;\big[\Phi^{(1)}_{ac}(k;\eta_a,s') g_{bc}(k;\,\eta_a,s')+ g_{ac}(k;\,\eta_a,s')\Phi^{(1)}_{cb}(k;s',\eta_a) \big],
\label{T1loop}
\eea where $\Phi^{(1)}_{ac}(k;\eta_a,\eta_b)$ is the 1-loop approximation of $\Phi_{ac}(k;\eta_a,\eta_b)$ given by eq. (\ref{Phi1loop}), which corresponds to 
the right panel of Fig. \ref{SigmaPhi}.
On the other hand,  as we show immediately below,  the exact equation for  $k\to+\infty$ has a similar  structure, and can be written as
 \bea
&&\partial_{\eta_a} \,P_{ab}^{eik}(k; \eta_a) = -\Omega_{ac} \,P_{cb}^{eik}(k; \eta_a)  -\Omega_{bc} \,P^{eik}_{ac}(k; \eta_a)\nonumber \\
&&\;\; +H_{\bf{ a}}^{eik}(k;\, \eta_a,\eta_{in}) P_{{\bf a}b}^{eik}(k; \eta_{in})  +H_{\bf{ b}}^{eik}(k;\, \eta_a,\eta_{in}) P_{a{\bf b}}^{eik}(k; \eta_{in})  
\nonumber\\
&&\;\;+ \int \, ds'\;\big[\Phi^{eik,L}_{ac}(k;\eta_a,s') G^{eik}_{bc}(k;\,\eta_a,s')+ G^{eik}_{ac}(k;\,\eta_a,s')\Phi^{eik,R}_{cb}(k;s',\eta_a) \big].
\label{Teik}
\eea  To show that this equation is satisfied in the large $k$ limit we need to compute 
\be
\int_{\eta_{in}}^{\eta} d s\;\lb\Phi_{ad}^{eik,L}(k;  \eta,s)
G_{bd}^{eik}(k;\eta,s)+
G_{ac}^{eik}(k;\eta,s) 
\Phi_{cd}^{eik,R}(k; s, \eta)\rb.
\ee 
The diagrams that contribute to the first term are shown in Fig. \ref{PHIG}, where  the thick lines  represent  renormalized propagators  ($G^{eik}$),  the black PS correspond to $P^{eik}$, and the thick dots stand for renormalized  vertexes $\Gamma_{dce}^{eik}$, which are given by eq.s~(\ref{propeik}), (\ref{PSeik}) and (\ref{renvert}), respectively. The computation is analogous to the one of the previous section (see \cite{AnselmiPietroni2012} for additional details), and the result is
\bea
&& \int_{\eta_{in}}^{\eta} d s\;\Phi_{ad}^{eik,L}(k;  \eta,s)
G_{bd}^{eik}(k;\eta,s)\nonumber\\
&&=  \frac{e^{\eta}}{C(\eta)}G^{eik}(k,\eta,\eta_{in})G^{eik}(k,\eta,\eta_{in})u_au_b P_{in}(k)\sum_{m=1}^{+\infty}\frac{ (k^2\sigma_v^2)^m\,{\cal{I}}^{2m-1}(\eta,\eta_{in})}{(m-1)!\,m!}m!\nonumber\\
&&= \frac{e^{\eta}}{C(\eta)}G^{eik}(k,\eta,\eta_{in})G^{eik}(k,\eta,\eta_{in})u_au_b P_{in}(k)e^{k^2\sigma_v^2{\cal{I}}^2(\eta,\eta_{in})}\left[ k^2\sigma_v^2{\cal{I}}^2(\eta,\eta_{in})\right]\nonumber\\
&& = u_au_b P_{in}(k) \frac{e^{\eta}}{C(\eta)} \left[k^2\sigma_v^2{\cal{I}}(\eta,\eta_{in})\right],
\eea 
where in the first equality the factors $(m-1)!\,m!$ in the denominator come after using the property of eq.~(\ref{proper}) for the time integrals corresponding to the  $m-1$ and $m$ vertex insertions that are respectively on the left and right of the hard PS, and the factor $m!$ takes into account all the possible ways of contracting the lines. 

Therefore, proceeding similarly for the contribution involving  $\Phi_{cd}^{eik,R}$, we have
\bea
&&\int_{\eta_{in}}^{\eta} d s\;\lb\Phi_{ad}^{eik,L}(k;  \eta,s)
G_{bd}^{eik}(k;\eta,s)+
G_{ac}^{eik}(k;\eta,s) 
\Phi_{cd}^{eik,R}(k; s, \eta) \rb\nonumber\\
&&=2 u_au_b P_{in}(k) \frac{e^{\eta}}{C(\eta)} \left[k^2\sigma_v^2{\cal{I}}(\eta,\eta_{in})\right].
\eea     
Then, it is a straightforward check to plug this result along with eq.s~(\ref{TRGL2}) and (\ref{propeik})  into eq.~(\ref{Teik}) and obtain eq.~(\ref{PSeik}) as a solution.

Let us now  summarize the arguments of \cite{AnselmiPietroni2012} that motivate the use of $\tilde\Phi_{ab}$ given in eq.~(\ref{phitilde}) as an interpolation between the small $k$ and the large $k$ regimes. We start by recalling that in the large $k$ limit the diagrams that contribute to $\int_{\eta_{in}}^{\eta} d s\;\Phi_{ad}^{L}(k;  \eta,s)G_{bd}(k;\eta,s)$ can be split into two terms, as shown in Fig.~\ref{PHIG}. In fact, diagram A gives
\bea
&& \frac{e^{\eta}}{C(\eta)}G^{eik}(k,\eta,\eta_{in})G^{eik}(k,\eta,\eta_{in})u_au_b P_{in}(k)\sum_{m=1}^{+\infty}\frac{ (k^2\sigma_v^2)^m\,{\cal{I}}^{2m-1}(\eta,\eta_{in})}{(m-1)!\,m!}(m-1)!\nonumber\\
&=& \frac{e^{\eta}}{C(\eta)}G^{eik}(k,\eta,\eta_{in})G^{eik}(k,\eta,\eta_{in})u_au_b P_{in}(k)\frac{1}{{\cal{I}}(\eta,\eta_{in})}\lb e^{k^2\sigma_v^2{\cal{I}}^2(\eta,\eta_{in})}-1\rb,\nonumber
\eea 
while diagram B contributes as
\bea
&& \frac{e^{\eta}}{C(\eta)}G^{eik}(k,\eta,\eta_{in})G^{eik}(k,\eta,\eta_{in})u_au_b P_{in}(k)\sum_{m=2}^{+\infty}\frac{ (k^2\sigma_v^2)^m\,{\cal{I}}^{(2m-1)}(\eta,\eta_{in})}{(m-1)!m!}(m-1)^2(m-2)!\nonumber\\
&=& \frac{e^{\eta}}{C(\eta)}G^{eik}(k,\eta,\eta_{in})G^{eik}(k,\eta,\eta_{in})u_au_b P_{in}(k)\frac{1}{{\cal{I}}(\eta,\eta_{in})}\left\{ 1-e^{k^2\sigma_v^2{\cal{I}}^2(\eta,\eta_{in})}\left[1-k^2\sigma_v^2{\cal{I}}^2(\eta,\eta_{in})\right]\right\}.\nonumber
\eea

Notice though that for the physically relevant scales, the large $k$ limit for the vertexes is not a good approximation and only in that limit the simplification in eq.~(\ref{renvert}) for the renormalized vertexes occurs, and the time integrations over the vertex insertions decouple from the ones correcting the propagators on the hard lines.  Therefore, as a first step towards a better approximation, one could  replace the rightmost vertexes in Fig.~\ref{PHIG} by tree-level ones. Doing this, the contribution to $\Phi$ given by diagram A looks more similar to its 1-loop counterpart  (which is more appropriate in the small $k$ regime), with the only difference that the hard PS  is taken to be the renormalized one, $P^{eik}$. However, for small values of $k$ this difference produces a correction of order $k^4$. Hence, to recover the appropriate behavior for $k\to 0$,  we need both to replace the rightmost vertexes by 
tree-level ones and to suppress the contribution of diagram B.  The latter can be  implemented by introducing the filter function $F(k)$ defined in eq.~(\ref{filter}).

\begin{figure}[t]
\begin{center}
\includegraphics[width=0.9\textwidth]{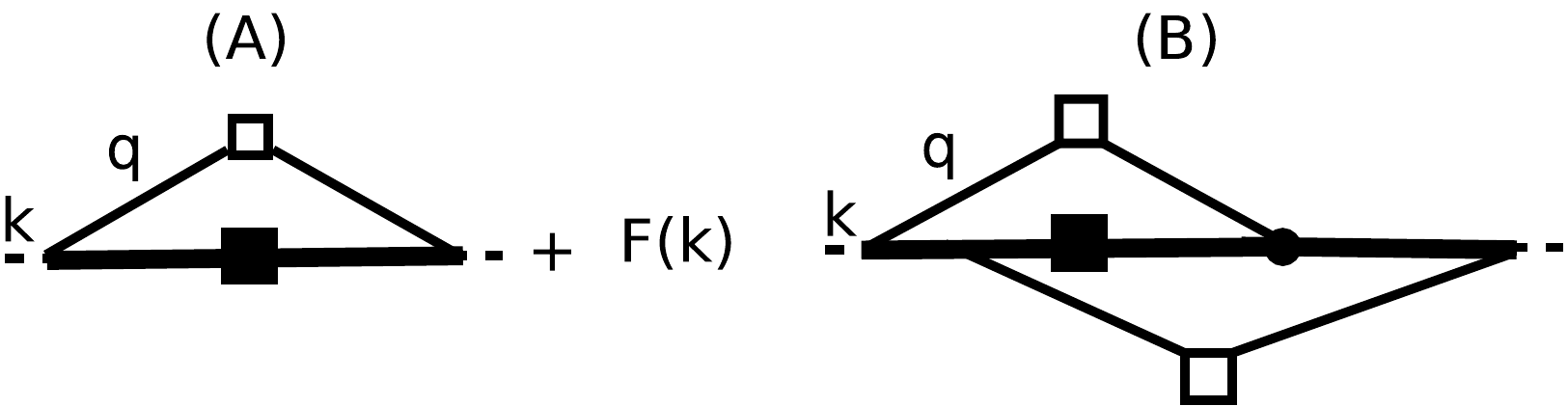}
\end{center}
\caption{  Diagrams contributing to $\tilde{\Phi}_{ab}$. }
\label{PHITilde}
\end{figure}

To sum up, the resulting approximation for $\Phi_{ab}(k,s,s')$ can be written as the weighted sum of the diagrams in Fig.~\ref{PHITilde}. The contribution of diagram A is given by
\be
 \exp\left[-\frac{1}{2}\,k^2\,\sigma_v^2\,{\cal{I}}^2(s,s')\right]\Phi_{ab}^{(1)}(k,s,s'),
\ee  
while the one of diagram B is
\bea
&&  \frac{e^{s+s'}u_au_b P_{in}(k)}{C(s)C(s')}G^{eik}(k,s,\eta_{in})G^{eik}(k,s',\eta_{in})\sum_{m=2}^{+\infty}\frac{ (k^2\sigma_v^2)^m\,\left[{\cal{I}}(s,\eta_{in}){\cal{I}}(s',\eta_{in})\right]^{m-1}}{(m-2)!}\nonumber\\
&&= \exp\left[-\frac{1}{2}k^2\sigma_v^2{\cal{I}}^2(s,s')\right]\frac{e^{s+s'}}{C(s)C(s')}\,(k^2\sigma_v^2)^2\,{\cal{I}}(s,\eta_{in})\,{\cal{I}}(s',\eta_{in})\,u_a\,u_b\, P_{in}(k)\,,
\eea
resulting in eq.~(\ref{phitilde}).

%%%%%%%%%%%%%%%%%%%%%%%%%%%%%%%%%%%%%%%%%%%%%%%%%%%%%%%%%%%%%%%%%%%%%%%%%%%%%%%%%%%%%%%%%%%%%%%%%
%%%%%%%%%%%%%%%%%%%%%%%%%%%%%%%%%%%%%%%%%%%%%%%%%%%%%%%%%%%%%%%%%%%%%%%%%%%%%%%%%%%%%%%%%%%%%%%%%
\section{Testing the AP resummation} 
\label{app:MICEcomp}

\begin{figure}[t]
\begin{center}
{\includegraphics[width=0.9\textwidth]{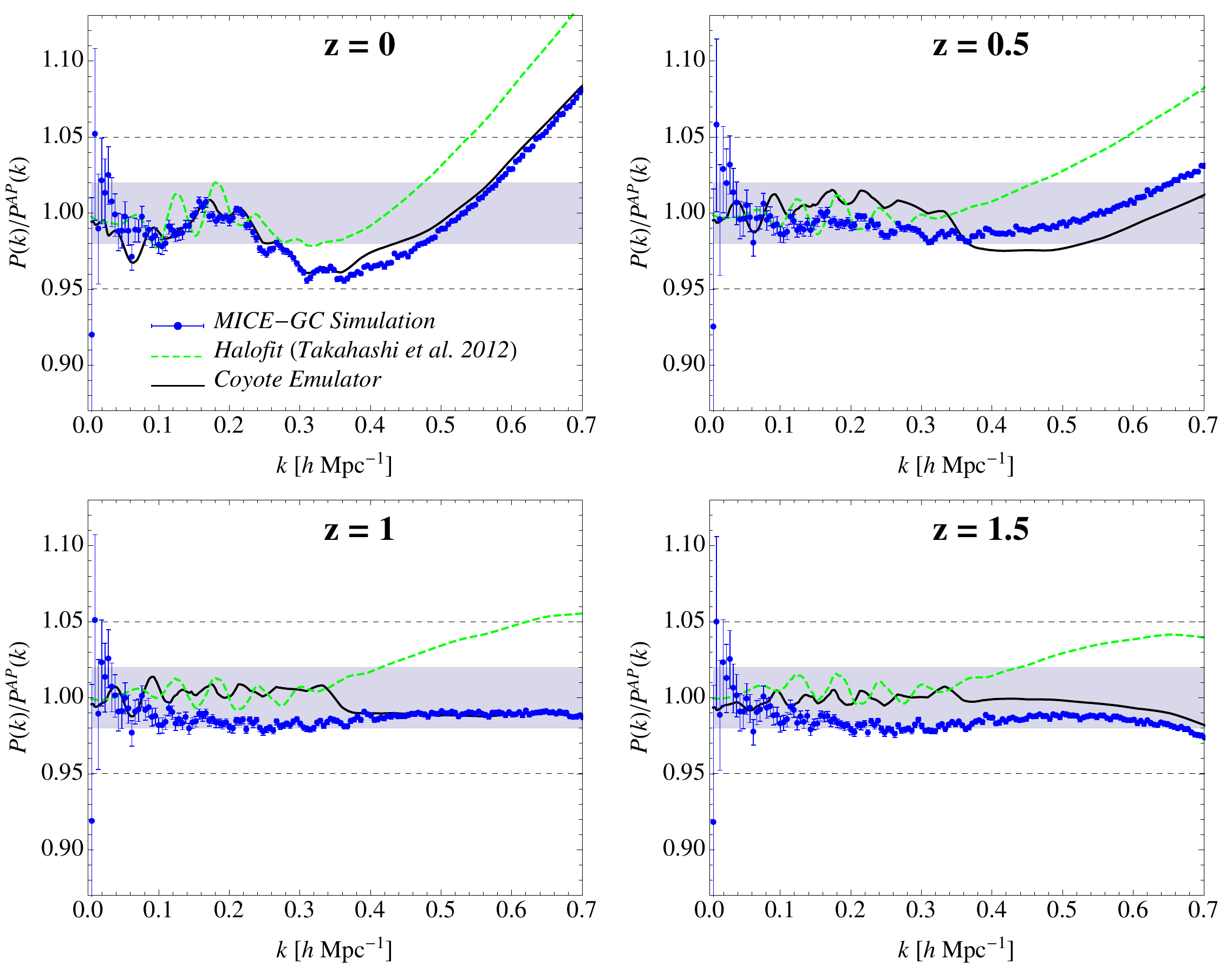}}
\caption{\small{The {\em blue dots} represent the ratio of the matter power spectrum measured in N-body simulations to the AP prediction at z=0 ({\em top left}), z=0.5 ({\em top right}), z=1 ({\em bottom left}) and z=1.5 ({\em bottom right}). Also shown are the outputs given by the revised  \texttt{halofit} \cite{TakahashiEtal2012} ({\em dashed-green line}) and the Coyote Emulator \cite{HeitmannEtal2014} ({\em black line})} codes.}
\label{fig:MICE}
\end{center}
\end{figure}

In this appendix we test the accuracy of the AP resummation comparing it to N-body simulations measurements as well as other predictions for the nonlinear power spectrum. Since there are not simulations for clustering quintessence cosmologies so far, we will limit ourselves to a test for a $\Lambda$CDM cosmology. In the original AP paper \cite{AnselmiPietroni2012} the authors made a careful comparison with numerical results in the BAO regime. Here we extend this test to smaller scales with the recent, high-resolution MICE-GC simulation \cite{FosalbaEtal2013A, CrocceEtal2013, FosalbaEtal2013B}.

Fig.~\ref{fig:MICE} shows the N-body data divided by the AP predictions ({\em blue dots with error bars}). For comparison, also shown are the Coyote Emulator prediction \cite{HeitmannEtal2014} ({\em black line}) and the revised  \texttt{halofit} formula of \cite{TakahashiEtal2012} ({\em green--dashed line}). Different panels correspond to different redshifts, that is $z=0$ ({\em top-left}), $z=0.5$ ({\em top-right}), $z=1$ ({\em bottom-left}) and $z=1.5$ ({\em bottom-right}). For all the redshifts considered, in the BAO range of scales, the overall matching simulations-AP is always better than $2\%$. At smaller scales, the agreement is still at the $2\%$ level for  $k \lesssim 0.6  h \text{ Mpc}^{-1}$ at $z=0.5$, and it extends to $k \lesssim 0.7  h \text{ Mpc}^{-1}$  for $z\gtrsim0.5$. At $z=0$ AP deviates up to $2\%$ for $k \lesssim 0.25  h \text{ Mpc}^{-1}$ and  up to $5\%$ for $k \lesssim 0.6  h \text{ Mpc}^{-1}$. As underlined in the main text, this is expected given the intrinsic limit of the PPF 
approximation.

The revised  \texttt{halofit} and the Extended Coyote Emulator results have been already compared with the MICE-GC measurements in
\cite{FosalbaEtal2013A}. We confirm that both codes work at $2\%$ in the BAO regime. This still holds for smaller scales for 
the Coyote Emulator while the revised  \texttt{halofit} performs at $\sim 5\%$ level beyond BAO.

\bibliographystyle{JHEPb}
%\bibliography{cosmologia}

\providecommand{\href}[2]{#2}\begingroup\raggedright\endgroup

\end{document}